\documentclass[aps,prb,twocolumn, amsmath, amssymb, superscriptaddress]{revtex4-2}
\usepackage{graphicx}
\usepackage{bm}
\usepackage{amsmath}
\usepackage{float}
\usepackage[FIGTOPCAP]{subfigure}
\usepackage[framemethod=tikz]{mdframed}
\usepackage{bbold}
\usepackage{amsmath}
\usepackage{amssymb}
\usepackage{dsfont}
\usepackage{physics}

\usepackage{hyperref}

\begin{document}

\title{Topological phase transition driven by structural defects}

\author{Andrii Syrota}
\affiliation{Universit\'e Paris-Saclay, CNRS, Laboratoire de Physique des Solides, 91405, Orsay, France}
\author{Andrej Mesaros}
\affiliation{Universit\'e Paris-Saclay, CNRS, Laboratoire de Physique des Solides, 91405, Orsay, France}
\author{Pascal Simon}
\affiliation{Universit\'e Paris-Saclay, CNRS, Laboratoire de Physique des Solides, 91405, Orsay, France}

\date{\today}

\begin{abstract}
Structural defects, such as disclinations and dislocations, destroy the long-range crystalline order as they proliferate. In this work, we continuously drive a system from a crystalline state on a decorated honeycomb lattice to a hyperuniform amorphous state, by consecutively introducing Stone-Wales (SW) defects, which can be viewed as dipoles of dislocations.
Using a topological Weaire–Thorpe Hamiltonian model to describe electrons in this system, 
we demonstrate that SW defects can cause pseudo-band inversions and, correspondingly, a reversal of the Chern number. Using adiabatic arguments and a computationally efficient spillage indicator, we predict the topological phase diagram of the amorphous state from the crystalline one.
We explain the nature of the pseudo-band inversion through the renormalization of the hopping in an effective Hamiltonian based on a single SW defect.
\end{abstract}

\maketitle

\section{Introduction}
Symmetry-protected topological phases of matter - like insulators and superconductors - have been studied in detail in crystalline systems, where the translational long-range order allows a development of topological band theory, leading to an exhaustive classification of crystalline systems in absence of electron interactions \cite{bradlyn2017topological, PointGroupInv, po2017symmetry, benalcazar2014classification}. Recent progress, both theoretical \cite{Agarwala, mansha2017robust, Marsal-amorphous-TQM, poyhonen2018amorphous, zhang2023anomalous, Corbae23} and experimental \cite{mitchell2018amorphous, zhou2020photonic, liu2020topological, corbae2023observation, ciocys2024establishing}, offers new insights into how amorphous materials, lacking periodic structure, could host nontrivial topology. The usual approach to study the topological properties of systems without translational symmetry, is based on the numerical computation of real-space topological indicators, such as the local Chern marker \cite{local-chern-marker}, the Bott index \cite{bott-index}, or the spectral localizer \cite{spectral-localiser}, which often require significant computational resources. Given the rising interest in topological amorphous matter due to its potentials and challenges\cite{adolfobook,Corbae23,Liu2025}, new insights and characterization methods demanding less resources would be highly valuable.

Some progress was made in developing computationally efficient analytical indices based on purely local symmetries \cite{Marsal-amorphous-TQM} or on global ensemble-averaged ones that are effectively present in the system \cite{statistical-topo-insulators}. In particular, the authors of \cite{Marsal-amorphous-TQM} introduced a class of modified Weaire–Thorpe (WT) models \cite{Weaire-1971} using a Voronoi construction from random points, so that the local coordination number of each site remains constant, while the graph of nearest-neighbor connections has no translational order. They then computed a Chern number by defining an analytical symmetry indicator, which relies on evaluating a local symmetry operator eigenvalue for momenta zero and infinity, in analogy with the orbital angular momentum invariant developed for a homogeneous gas in \cite{homogeneus-gas-invariant}. The question remains if more general methods could be found, and if the phase diagram has a simple interpretation.

In this work, we approach the same WT model from the opposite limit:
We explain the emergence of topological phases in a wide class of amorphous graphs by deriving them from the underlying crystalline topological phases on the honeycomb lattice. For that we use a controlled approach (Sec. \ref{sec:model-hamiltonian}) to continuously go from the crystalline lattice to an amorphous graph by introducing SW topological defects \cite{stone-wales}. The process allows us to control the \textit{local} distribution of plaquette (loop) sizes, allowing various amorphous systems starting from the same crystalline model.

In the initial crystalline model, we first show that the topological gaps (computed by crystalline symmetry indicators \cite{PointGroupInv}) can be predominantly traced back to the $\mathbf{K}$-point of the Brillouin zone, whose existence is tightly bound to the crystalline structure. The adiabatically added structural disorder hence changes the topology of the system at the filling factors that correspond to the gaps labeled by the $\mathbf{K}$-point.
 
We introduce the quantity of eigenstate spillage, as a computationally cheap measure of overlap between crystalline and amorphous states, revealing that in certain regions of the phase diagram eigenstates near the gap retain their extended crystalline nature. We can hence apply the notion of band inversion in crystalline systems even to the situation where the system passes through a phase transition due to becoming increasingly amorphous. This allows us to extend the formula of original crystalline symmetry indicators into the amorphous ground state, and partially predict the topological phase diagram of amorphous systems by only knowing the topological  phase diagram of the associated crystal plus the knowledge of the two gap-edge eigenstates of the amorphous system. We then verify our phase diagram with the numerical calculation of the Bott index.
 
Finally, we give a physical picture of the topological transitions that occur with amorphization, by considering a configurational average over a single SW defect. We demonstrate that this topological defect effectively introduces sublattice-valley polarization for intra-atom hopping terms of the Bloch Hamiltonian, which causes the shift of the original crystalline bands, and therefore the effective band inversion.
 
Our plan of the paper is as follows: 
In  Sec. \ref{sec:model-hamiltonian}, we introduce our procedure to continuously go from a crystalline lattice to an amorphous one together with our electronic model Hamiltonian. Next, in  Sec. \ref{sec:ele-topo-properties}, we analyze in detail the electronic and topological properties of the system. Then, in  Sec. \ref{sec:topology} we study the adiabatic connection between the  crystalline and amorphous topology using the spillage-assisted indicator that we introduce. In Sec. \ref{sec:sigma}
we derive an effective model, whose renormalized intra-atom hopping drives the topological transition. Finally in Sec. \ref{sec:conclusions}, we conclude with some discussion.

\section{Amorphization procedure and model Hamiltonian}
\label{sec:model-hamiltonian}
\begin{figure}[h!]
\includegraphics[width=0.5\textwidth]{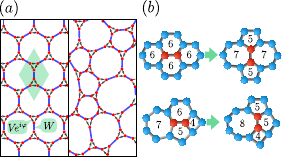}
\caption{(a) Comparison between crystalline (left) and amorphous (right) lattices of the three-fold coordinated W-T model with broken time-reversal symmetry. Blue edges, red dots, and black oriented triangles correspond to inter-atom hopping, orbitals' positions, and flux-enhanced intra-atom hopping, respectively. Directions of arrows depict the positive phase gain of the electrons due to magnetic flux. (b) Example of Stone-Wales defect, which changes the sizes of neighboring plaquettes. The upper figure shows how a Stone-Wales defect affects the crystalline lattice, while the lower one shows that consecutive application of defects can yield plaquettes of any size $\geq3$.}
\label{fig:models}
\end{figure}

In order to present our amorphization procedure we start with a 2D 
honeycomb lattice (Fig.~\ref{fig:models}(a)) with periodic boundary conditions (PBC) and consecutively insert SW defects. To do so, we  (1) randomly select a bond; (2) move the two atoms that form it by rotating the bond around its mid-point by $\pm\pi/2$, with randomly chosen sign; (3) reconnect the 4 remaining bonds that emanate from the two atoms (see Fig.~\ref{fig:models}(b), upper). Importantly, this procedure keeps the local three-fold coordination of each atom. Hence, introducing a defect is well defined on the systems graph, no matter the number or position of defects already introduced (Fig.~\ref{fig:models}(b), lower). The procedure of introducing SW defects preserves the total number of bonds and the total number of plaquettes. We choose to preserve the \textit{original labeling of the sites}, as discussed in Section~\ref{sec:topology}. A SW defect introduced into a pristine honeycomb part of the system locally changes the configuration of the plaquette sizes: $(6,6,6,6)\rightarrow (7,7,5,5)$. Importantly, adding further defects allows the formation of plaquettes of any size (see Fig.~\ref{fig:models}(b), lower). The  whole amorphization procedure is detailed in Appendix~\ref{app:Appendix-lattice-amorphisation}.

We define an amorphization parameter $\alpha$ as:
\[
\alpha = \frac{N_{non-hex}}{N_{total}},
\]
where $N_{non-hex}$ is the number of non-hexagonal plaquettes, and $N_{total}$ is the total (conserved) number of plaquettes in the system. Therefore, $\alpha = 0$ means that the system is in the honeycomb crystalline configuration and $\alpha = 1$ means that SW defects proliferated, producing an amorphous state with no hexagonal plaquettes left. In order to have a monotonous amorphization, we choose to reject a SW defect (step (1) of the procedure) if it increases the number of hexagonal plaquettes, i.e., if it would decrease $\alpha$.

To study the electronic properties of the system, we use the Hamiltonian originally introduced by Weaire and Thorpe \cite{Weaire-1971} and add the magnetic flux that breaks time-reversal symmetry (TRS) following Ref. \cite{Marsal-amorphous-TQM}. The key feature of this tight-binding model is that it suits any random lattice with fixed local coordination (in our case three-fold). The Hamiltonian consists of two types of terms. The first one is an intra-site hopping (arrowed triangles in Fig.~\ref{fig:models}(a)) between the three orbitals  (red dots) of an atom with complex hopping magnitude $Ve^{i\varphi}$, incorporating the magnetic flux $\varphi$. The second type of terms is inter-site hopping (blue lines in Fig.~\ref{fig:models}(a)), with magnitude $W$, which is kept constant regardless of the real-space length of the bond between atoms. For simplicity, hereafter we set $V=1-W$. The full Hamiltonian is then written as:
\begin{equation}
\label{eq:Waire Thrope Hamiltonian}
\hat{\mathcal{H}}=\hat{\mathcal{H}}_V+\hat{\mathcal{H}}_W=\sum_{i, j \neq j^{\prime}} V_{j j^{\prime}} \hat{c}_{i j}^{\dagger} \hat{c}_{i j^{\prime}}+\sum_{i \neq i^{\prime}, j} W \hat{c}_{i j}^{\dagger} \hat{c}_{i^{\prime} j},  
\end{equation}
where $\hat{c}_{ij}$ annihilates an electron in orbital $j=1,2,3$ of atom labeled by $i$. The $V_{j j^{\prime}} $ terms are given by the matrix
 \begin{equation}
\label{eq:Vmatrix} 
V_{j j^{\prime}}=
\begin{bmatrix}
0 & V e^{i \varphi} & V e^{-i \varphi}  \\
V e^{-i \varphi} & 0 & V e^{i \varphi}  \\
V e^{i \varphi} & V e^{-i \varphi} & 0  \\
\end{bmatrix}_{j,j'},
\end{equation}
describing the hopping between the different orbitals of the same atom. The sign of the magnetic flux is given by the hopping direction: the flux is $+\varphi$ if the electron is hopping clockwise on the triangle of orbitals, and $-\varphi$ if electron hops anti-clockwise. Note that SW defects do not alter the order of orbitals in an atom. We fix the value $V>0$, and the magnitude $|\phi|$, to be identical for all atoms regardless of their position.

\section{Electronic and topological properties}
\label{sec:ele-topo-properties}
As was pointed out originally by  Weaire and Thorpe \cite{Weaire-1971}, the electronic properties of solids are strongly dominated by their short-range order. Therefore, one could infer the electronic properties of certain types of amorphous systems by analyzing crystalline systems with an identical local environment. In comparison to the resolvent method suggested in Ref. \cite{Schwartz-resolvent} and used more recently in \cite{Marsal-amorphous-TQM} to analytically determine the key regions with non-zero density of states (DOS), here we rely on a controlled amorphization procedure in order to use the crystalline model to reveal topological and spectral properties in the associated amorphous system. We hence first analyze the crystal model, which will be amorphized with SW defects.

\subsection{Electronic properties of the crystalline lattice}
\label{sec:electronic_properties}
The crystalline model, presented in Fig.~\ref{fig:models}(a), is a honeycomb lattice, with a unit cell (green rhombus) which incorporates six orbitals, effectively forming two sublattices, each containing three orbitals. We make use of the symmetry indicator (SI) method to compute the Chern number associated to certain insulating filling factors, $f=n/6,~n\in 1...5$.
In our case, the lattice has the symmetry of the $p6$ wallpaper group, and therefore carries a $\mathbb{Z}_6$ topological invariant\cite{po2017symmetry}, requiring us to consider the symmetry representations of the operators of six-fold ($\hat{R}_6$), three-fold ($\hat{R}_3$), and two-fold ($\hat{R}_2$) rotations, at the $\mathbf{\Gamma}$, $\mathbf{K}$, and $\mathbf{M}$ point of the Brillouin Zone (BZ), respectively. The topological symmetry indicator for the Chern number is then computed as:
\begin{equation}
\label{eq:symm-ind}
Ch_{\text{ind}} =\frac{3}{i\pi}\log\left[\prod_{i \in o c c .} \eta_i(\mathbf{\Gamma}) \theta_i(\mathbf{K}) \zeta_i(\mathbf{M})\right] ~\text{mod} ~6,
\end{equation}
where the product is taken over all the occupied Bloch bands labeled by $i$, while $\eta_i, \theta_i,$ and $\zeta_i$ are $i$-th state's eigenvalue of $\hat{R}_6$, $\hat{R}_3$, and $\hat{R}_2$, respectively. 

\begin{figure*}[hbt!]
\includegraphics[width=1\textwidth]{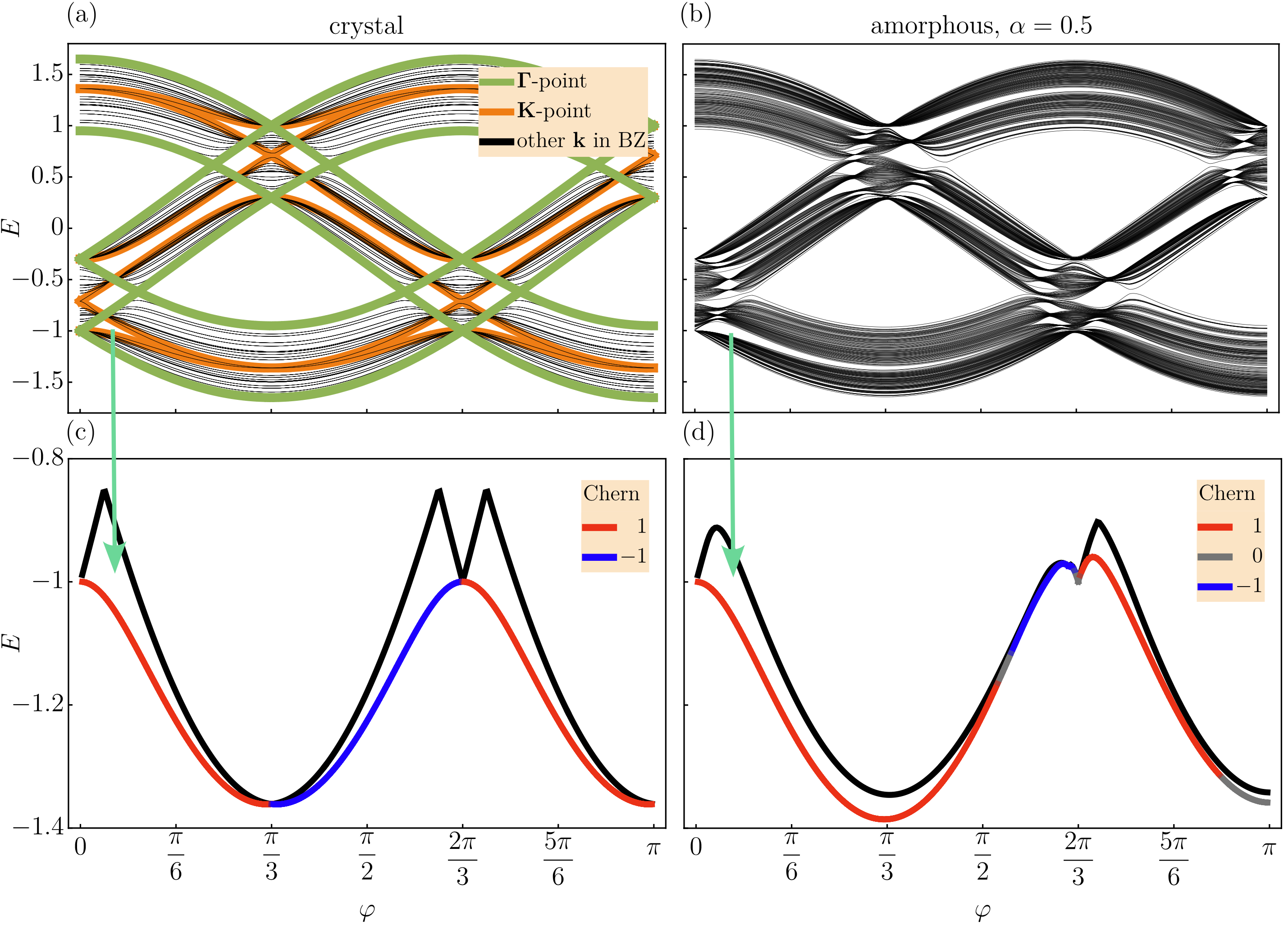}
\caption{Energy spectrum (solid black lines) as function of $\varphi$ at fixed $V=0.65$, and for $\alpha=0.5$, in the crystalline (a), and amorphous (b) system. System size is chosen to be small: 81 unit cells (u.c.). in order not to overwhelm the plots graphically. (a) $\mathbf{\Gamma}$-states (solid green lines), and $\mathbf{K}$-states (solid orange lines) reveal the presence of mini-gaps within the global spectral bands, which are the only topological gaps in the system. (c) and (d) show only the two lowest-energy eigenstates forming the mini-gap at filling factor 1/6 (shown explicitly by green arrows). The color of the lower eigenstate corresponds to the Chern number of the mini-gap computed by the Bott index method. The crossing point which occurs in $\varphi=\pi/3$ in (c) and is shifted to a larger value of $\varphi$ in (d) corresponds to a Chern number inversion from -1 to 1 in that range of $\varphi$.}
\label{fig:cry-amo-spectra}
\end{figure*}

We present the DOS dependence on $\varphi$ in Fig.~\ref{fig:cry-amo-spectra}(a), labeling gaps by their Chern numbers computed by the SI method. First, the global gap edges of the DOS are given by states at the $\Gamma$-point, in agreement with the prediction of the resolvent method. Second, the gap edges of mini-gaps are predominantly given by eigenstates at the $\mathbf{K}$-point. Moreover, the non-trivial topological states occur at fillings which are inside the mini-gaps, while the global gaps are topologically trivial.

\subsection{Electronic properties in the amorphous lattice}
One can generally expect that $\mathbf{\Gamma}$-states are more robust than $\mathbf{K}$-states to local lattice perturbations, such as SW defects. Therefore, the strongest evolution of a topological phase diagram under amorphization should happen in mini-gaps associated with the $\mathbf{K}$-point. We henceforth fix the filling factor $f$ to be $1/6$. Fig.~\ref{fig:cry-amo-spectra}(b) illustrates that the main effect of amorphization in the spectrum is indeed a strong modification of mini-gaps, while the global gaps remain intact. 

We identify two main types of behavior of a spectral mini-gap: (1) it remains open under amorphization, with its boundaries possibly shifting, or (2) it becomes completely filled with impurity states and hence closes. In this paper, we will focus on the first scenario, and the associated topological phase transition, while the second scenario will be discussed elsewhere.

To compute the Chern number in the amorphous case, we use the Bott index \cite{bott-index}, which has been proven to coincide with the Chern number in the thermodynamic limit for systems with PBC. To implement it, we first construct the mapping from real-space positions of the atoms to angular coordinates defined on the torus:
\begin{equation}
    (x,~y)\mapsto(\theta_x = e^{x\frac{2\pi i}{L_x}},~\theta_y = e^{y\frac{2\pi i}{L_y}}),
\end{equation}
where $L_x,~$and $L_y$ are linear dimensions of the system. We set the spatial positions of the orbitals of the same atom to be equal. The usual Bott index is then:
\begin{equation}
\label{eq:BottIndex}
\text{Bott} = \frac{1}{2\pi} \operatorname{Im}\left\{\operatorname{Tr}\left[\log \left(U_x U_y U_x^{\dagger} U_y^{\dagger}\right)\right]\right\},
\end{equation}
where the almost-unitary matrices $U_x,U_y$ are:
\begin{equation}
U_x=\mathcal{P}^\dagger\Theta_x \mathcal{P},\quad \quad
U_y=\mathcal{P}^\dagger\Theta_y \mathcal{P},
\end{equation}
with $\Theta_{x/y}$ the diagonal matrices constructed from values of $\theta_{x/y}$ for corresponding orbitals, while $\mathcal{P}= \left[ |\psi_{ 1}\rangle,~|\psi_{ 2}\rangle,~...~ |\psi_{ n}\rangle\right]$ are matrices constructed from the occupied subspace of the Hamiltonian \cite{loring2019guide} (n is a number of states below Fermi level). 

To demonstrate the evolution of the mini gap, in Fig.~\ref{fig:cry-amo-spectra}(c,d) we focus on the two lowest energy eigenstates encompassing the mini gap corresponding to the filling $f=1/6$, and we compare the crystal to the amorphized lattice at $\alpha=0.5$. The Chern number in the mini gap evolves under the amorphization, as $\alpha$ changes:
in the region $0\leq\varphi\leq 2\pi/3$, we identify the single crossing point $\varphi_{c}(\alpha)$ between the two lowest-energy eigenstates. Its value shifts from $\varphi_{c,\textrm{cry}}\equiv\varphi_c(\alpha=0) = \pi/3$ in the crystal to a higher value $\varphi_{c}(\alpha>0)$ in the amorphous system. Consequently, the Chern number inverts in the region of the shift, i.e., in the region $\varphi_{c,\textrm{cry}}<\varphi<\varphi_{c}(\alpha>0)$ (the nature of the shift is discussed in Sec. \ref{sec:sigma}). This behavior suggests an interpretation in terms of a \textit{band-inversion}, even though in the amorphous  case the electronic bands are not properly well defined. The fact that the spectral mini-gap is open throughout the amorphization process (except at the single value $\varphi=\varphi_c(\alpha)$), implies an adiabatic connection between the topologies of the crystalline and the amorphous systems, which is the main focus of the next section.

\section{Adiabatic connection of crystalline and amorphous topology: spillage-assisted indicator}
\label{sec:topology}
Here we present a main result of this work, namely the way to adiabatically connect the crystalline and amorphous topologies. We rely on the assumption that a gapped ground state stays in the same topological phase unless a quantum phase transition occurs by closing the mobility gap. The mobility gap is relevant here because we have in mind the physical observable of a quantized Hall conductivity, associated with the Chern number. Therefore, the value of the Chern number assigned to a filling inside a mini-gap in the crystal should be also assigned to the amorphous ground states that we obtain by adiabatically tuning $\alpha$, as long as the mobility gap does not close at that filling. To measure the mobility gap, we introduce the quantity $\gamma$, inspired by eigenstate spillage, which was itself based on the idea of signaling band inversion in crystals due to spin orbit coupling\cite{spin-orbit-Spillage}, and was extended to non-crystalline systems with the use of plane-wave decomposition\cite{struct-spillage}.

\begin{figure}[hbt!]
\includegraphics[width=0.49\textwidth]{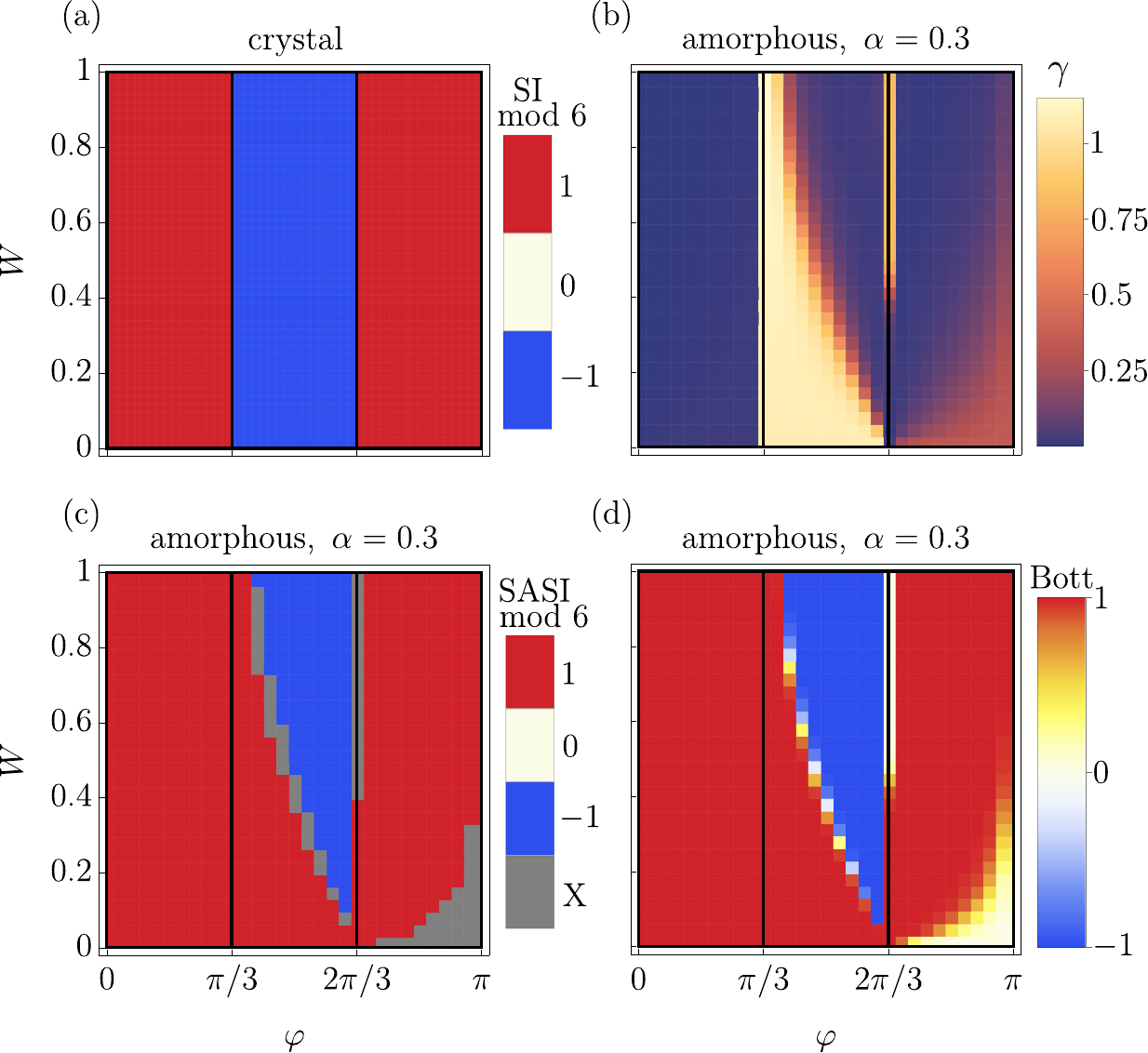}
\caption{Topological phase diagrams at 1/6 filling for system size of 144 u.c. (a) Chern number computed by the symmetry indicator method (Eq. \ref{eq:symm-ind}) for the crystalline system. (b) Eigenstate spillage computed by Eq. \ref{equation:MySpillage-full}, (c) Chern number computed by the by SASI with the following chosen cutoffs: $\gamma<0.25$ - no transition occured, $0.25<\gamma<0.7$ - unknown topological phase (gray region X), and $0.7<\gamma$ - topological phase transition given by Eq .\ref{eq:SASI}.(d) Chern number computed by Bott index (Eq. \ref{eq:BottIndex}). For the amorphous system we take $\alpha=0.3$ and average over 200 disorder realizations. }
\label{fig:main-result}
\end{figure}

Our spillage indicator $\gamma$ includes simply the two lowest-energy eigenstates around the minigap at a given filling, i.e., (1) the highest occupied eigenstate, which is the lower edge of the mini-gap, and (2) the lowest unoccupied eigenstate, which is the upper edge of the mini-gap, exactly as illustrated in Fig.~\ref{fig:cry-amo-spectra} (c,d). We compute the overlap between these eigenstates in the two systems, labeling by "cry" the crystal system, and by "amo" the amorphous system at the chosen $\alpha>0$:
\begin{equation}
\label{equation:MySpillage-full}
\gamma = \sum_{i \in deg} |\bra{\psi^{\text{occ}}_{\text{cry},~i}}\ket{\psi^{\text{empty}}_{\text{amo}}}|^2 + |\bra{\psi^{\text{empty}}_{\text{cry},~i}}\ket{\psi^{\text{occ}}_{\text{amo}}}|^2,
\end{equation}
where the sum is taken over the (possibly) degenerate subspace of a crystal (e.g., degeneracy due to two $\mathbf{K}$-points), and the second term is added to take into the account that the highest occupied eigenstate in an amorphous system may be preserved better under amorphization than the lowest empty one.

To be able to reasonably compare the eigenstates as $\alpha$  adiabatically changes, it is important that our procedure preserves as closely as possible the labeling of the sites at each step of the amorphization, i.e.,  each time a SW defect is introduced. Namely, there is an inherent  ambiguity in the labeling of the two atoms in the core of the SW defect, as there is no physical difference between rotating the bond clock- or anticlock-wise (we choose to systematically rotate clockwise). In the Appendix \ref{app:Appendix-spillage-errors}, we show that for the range of $\alpha$ considered here, the ambiguities remain local and observables such as the spillage indicator are essentially independent of the choice made.

Physically, the spillage $\gamma$ becomes non-zero when a pseudo-band inversion occurs. We present the disorder-averaged spillage, computed by Eq.~\eqref{equation:MySpillage-full} for various values of $W$, in Fig.~\ref{fig:main-result}(d). We identify three regions in the phase diagram, with distinct values of $\gamma$:
\begin{enumerate}
\item The spillage is low, hence no band inversion nor hybridization occurred between the two eigenstates, implying the same topological phase in the amorphous and in the crystal system ($\gamma<0.25$ in Fig. \ref{fig:main-result}(b)).
\item The spillage is high, hence pseudo-bands were inverted, while the two eigenstates preserved their crystalline-like nature to a large degree ($0.7<\gamma$ in Fig. \ref{fig:main-result}(b)). 
\item The spillage is intermediate, implying a strong hybridization of the two eigenstates; however, in this region there is a closure of the mobility gap without any reopening (not even in the limit of high $\alpha$), and therefore the amorphization destroys the (topological) gapped phase ($0.25<\gamma<0.7$ in Fig. \ref{fig:main-result} (b)).
\end{enumerate} 

The spillage therefore allows us to effectively identify the parts of the phase diagram, where the amorphization of the mini-gap could be interpreted in terms of no topological change  (case 1 above), or a potential change of topology due to a band-inversion-like event (case 2 above). By performing an analysis of the error estimation (detailed in Appendix \ref{app:Appendix-spillage-errors}), we determined that phases can be robustly discriminated for up to the  disorder strength $\alpha=0.35$ with a confidence level of 95\%. Notice that this can also serve as an indicator of the range over which amorphous states start to significantly lose their crystalline signature. Consequently, for the  case 2 indicating a band inversion, we now define a spillage-assisted symmetry indicator (SASI) for the Chern number of an amorphous system which takes the form:
\begin{equation}
\label{eq:SASI}
Ch^{\text{amo}}_{\text{ind}}\equiv Ch^{\text{cry}}_{\text{ind}} + \frac{3}{i\pi}\log \frac{ \theta_{\textrm{empty}}(\mathbf{K})}{\theta_{\textrm{occ}}(\mathbf{K})} ~\text{mod} ~6,
\end{equation}
which diagnoses whether the Chern number changes or not.
In Eq. \ref{eq:SASI}, $Ch^{\text{cry}}_{\text{ind}}$ is the crystalline SI (the full analytical computation of crystalline band symmetry indicators is detailed in Appendix \ref{app:Appendix-analytics}) while the second term indicates the change of the Chern number that would occur in the crystal if the two eigenstates have indeed inverted their places. More precisely, the formula is written in our example at filling $1/6$, for $\varphi$ values where the spillage indicates a band inversion. In that mini-gap the two lowest-energy eigenstates are at the $\mathbf{K}$-point of the crystal, so Eq.~\eqref{eq:SASI} simply takes into account the change of the $C_3$ symmetry eigenvalue that would occur in the crystal if the two eigenstates invert their place. The formula can thus be straightforwardly modified for other mini-gaps. In summary, the adiabaticity of the amorphization allows us to extend the symmetry indicator approach from the crystalline state given by Eq.~\eqref{eq:symm-ind} into the amorphous state.\footnote{We note that all the way up to $\alpha\leq 0.3$, the amorphous eigenstates remain visually  identifiable with the crystalline ones.}

In the parts of the SASI phase diagram where the spillage is large but the eigenstates bounding the mini-gap are formed by a mixture of different high-symmetry points, as this occurs  for example in the region close to $\varphi = 2\pi/3$ in Fig.~\ref{fig:main-result}(d), the formula is inapplicable. In this case, the mini-gap trivially closes, while mixing the occupied and empty state that  are, in the crystal, at $\mathbf{\Gamma}$ and $\mathbf{K}$-points, respectively.

Fig.~\ref{fig:main-result}(c) presents the topological phase diagram computed for an amorphous system using the SASI in Eq.~\eqref{eq:SASI}. The gray region of the diagram indicates where the formula is inapplicable. To benchmark our approach, we show the amorphous phase diagram computed by help of the Bott index in Fig.~\ref{fig:main-result}(b), and we also present the crystalline phase diagram computed with the SI in Fig.~\ref{fig:main-result}(a). Generally, the Bott index and the SASI results are in a good agreement, specifically in the regions where the Chern number inversion occurs. This indicates that using the amorphization procedure and the adiabatic argument gives a good partial prediction of the topological phase diagram of the amorphous system using the SASI.

The key point is that the computation of the SASI requires only a small number of low-energy eigenstates of the amorphous system, and the known topological phase diagram of the crystalline system. This enables a huge computation time-benefit, especially for very large systems, where the full diagonalization is not accessible, while one can easily obtain a few eigenstates (for example, by the Lanczos method).

\section{Renormalized intra-atom hopping drives the band inversion} 
\label{sec:sigma}

In this section, we explain the mechanism that drives, with growing $\alpha$, the shift of the crossing point of the $\mathbf{K}$-point states, and hence causes a pseudo-band inversion at a fixed $\varphi$, as presented in Fig.~\ref{fig:cry-amo-spectra}(d). We take the approach of an effective-medium self-energy \cite{sheng-disordered-wave-scattering} similarly to that used in \cite{topo-anderson-insulator}. This approach is valid in the regime where the system is not strongly localized under disorder, and the eigenstates remain extended, retaining crystalline-like signatures. This scenario is fulfilled in the region of our interest $0\leq\varphi\leq2\pi/3$, given by a large overlap between crystalline and amorphous eigenstates (Fig.~\ref{fig:main-result}(b)).

The idea behind this approach is to introduce an averaging over defect configurations (average T-matrix approximation \cite{economou2006green}) that would effectively restore the translational invariance of the self-energy:

\begin{equation}
\label{eq:self-ene}
    \Sigma = \left<T\right>_{c} (\mathbb{1}+G_0\left<T\right>_{c})^{-1},
\end{equation}
where $G_0 = (\varepsilon_F + i\eta-\mathcal{H}_{\text{cry}})^{-1}$ is the retarded Green's function of the crystal, with the Fermi energy $\varepsilon_F $ chosen to be in the middle of the mini-gap of interest. The $\eta$ is a small positive real number to ensure convergence, $\left<...\right>_{c}$ stands for configurational averaging, and the $T$-matrix is defined as:
\begin{equation}
\label{eq:T-matrix}
T = (\mathcal{H}_{\text{def}} - \mathcal{H}_{\text{cry}})(\mathbb{1}-G_0(\mathcal{H}_{\text{def}} - \mathcal{H}_{\text{cry}}))^{-1},
\end{equation}
where $\mathcal{H}_{\text{def}}$ is the Hamiltonian of the system with defects.
We now use the plain wave basis $\ket{\mathbf{k}} = \frac{1}{\sqrt{N}} \Sigma_l e^{i \mathbf{k}\cdot\mathbf{r}_l} \ket{l}$ of the original crystal to transform the computed self-energy to  momentum space, and then treat the hermitized self-energy as a perturbation to the Bloch Hamiltonian:
\begin{equation}
    \label{eq:Heff-amo}
    \mathcal{H}_{\text{eff}}(\mathbf{k}) = \mathcal{H}_{\text{cry}}(\mathbf{k}) + \Sigma(\mathbf{k}),
\end{equation}
where 
\[
\Sigma(\mathbf{k}) = \bra{\mathbf{k}} (\Sigma + \Sigma^{\dagger})/2 \ket{\mathbf{k}}.
\]

Topological lattice defects are an example of a correlated disorder that typically significantly couples the states with different quasi-momenta. Disorder averaging, however, somewhat reduces the effect of off-diagonal couplings due to effectively restoring the translational invariance. Even further, we examine the limit of very dilute disorder, making the off-diagonal terms of self-energy \textit{exactly zero} in momentum space. More precisely, we perform the configurational averaging by applying a single SW defect to all possible $W$-bonds present in the system. We also average over the two possible rotations of the bond for a given SW defect. This gives us $3\times2\times N_{\text{u.c.}}$ disorder configurations in total, for $N_{\text{u.c.}}$ unit-cells. 

\begin{figure}[hbt!]
\includegraphics[width=0.5\textwidth]{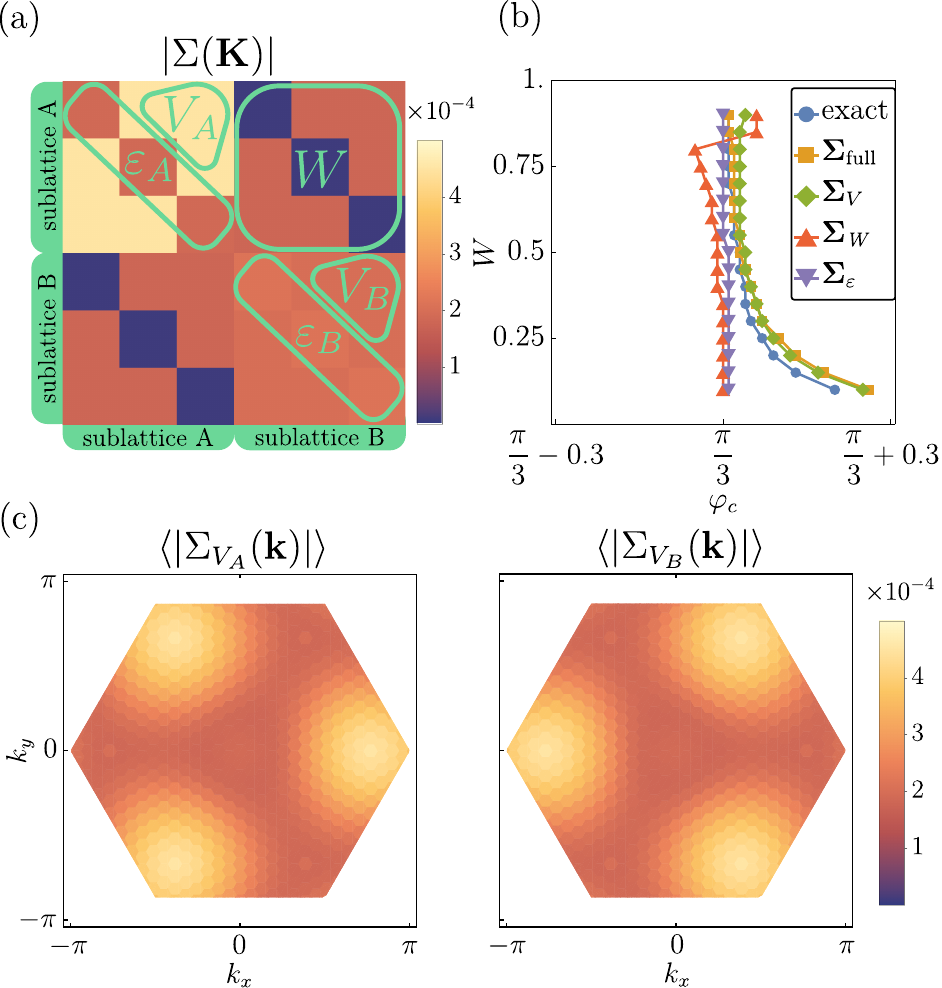}
\caption{(a) Self-energy matrix evaluated from Eq.~\ref{eq:self-ene} at the $\mathbf{K}$-point. Color intensity indicates the absolute value of each matrix element, showing that the dominant elements correspond to $V$-hoppings on the A sublattice. (b) The evolution of the crossing point $\varphi_c$ with $W$, due to one SW defect applied to the lattice. Exact diagonalization (blue) is compared to various contributions in the effective self-energy description. The largest shift of $\varphi_c$ is given by the $V$-terms of $\Sigma$ (green). (c) Distributions for separate components of $V$-terms of self-energy for sublattice A (left), and sublattice B (right) demonstrating signficant polarization close to each valley. (a) and (c) were computed for fixed values of parameters $V=0.8$, $\varphi=\pi/3+0.05$, $\eta = 0.003$, and system size of 324 u.c.. (b) was computed for varied range of parameters $V$, $\varphi$, but fixed $\eta = 0.003$, and system size of 144 u.c..}
\label{fig:sigma}
\end{figure}

Since all momenta are decoupled, it suffices to track the evolution of the crossing point $\varphi_c$ in the Hamiltonian in Eq.~\ref{eq:Heff-amo} only at the $\mathbf{K}$-point. First, we present the evaluated self-energy matrix in Fig.~\ref{fig:sigma}(a). It is a dense matrix: The terms on the main diagonal correspond to effective on-site energies $\varepsilon$ ($\varepsilon=0$ in original WT Hamiltonian), the terms given by the two diagonal $3\times 3$ blocks, without the main diagonal, are associated to the $V$-hopping between orbitals within the same sublattice, while the two off-diagonal $3\times 3$ blocks are associated with hopping between the two sublattices and correspond to $W$-terms of the initial model. Importantly, as shown in Fig.~\ref{fig:sigma}(a), the dominant entries in the matrix are the $V$-terms associated only with sublattice A. For the $\mathbf{K'}$-point however the dominant terms are that of sublattice B. In  Fig.~\ref{fig:sigma}(c) we show the mean of absolute values of $\Sigma(\mathbf{k})$ over the entire BZ for subllatice A (left), and sublattice B (right), which shows significant sublattice-valley polarization for the effective hopping terms. Polarization of other components of $\Sigma(\mathbf{k})$ close to each valley is negligible as discussed in greater detail in Appendix \ref{app:Sigma-details}.

Now we consider how each of the self-energy components affects the crossing point $\varphi_c$ for various $W$, see Fig.~\ref{fig:sigma}(b), and how it compares with the reference evolution (blue line) computed by exact diagonalization of a system with one SW defect. First, we find that the total effect of $\Sigma(\mathbf{k})$ (orange line) is in satisfactory agreement with the exact result (blue). Second, we confirm that the dominant contribution to the shift in $\varphi_c$ is indeed due to the $V$-terms in $\Sigma(\mathbf{k})$ (as it almost coincides with result for full $\Sigma(\mathbf{k})$), confirming the expectation from Fig.\ref{fig:sigma}(a).

We therefore arrive at the following physical interpretation: at low-enough $\alpha$, the shift of the crossing point occurs primarily due to the fact that the SW defect effectively renormalizes the intra-atom hopping for one of the sublattices at a given momentum. We note that this is not the case for the Anderson topological insulator, where the dominant effect is the renormalized topological mass \cite{topo-anderson-insulator}.  

The effective self energy also allows to probe the geometrical properties of the Bloch eigenstates, as it is well-defined for all allowed momenta in BZ. To confirm  the shift of the crossing point at $\varphi_c$, and also the Chern number inversion from -1 to 1, we have computed the Berry curvature in a gauge-invariant way, and have shown that it is indeed inverted under application of $\Sigma(\mathbf{k})$ close to each valley. Then, we also note that the $\Sigma(\mathbf{k})$ obeys all the point group symmetries of the crystalline Hamiltonian. This means that eigenstates of $\mathcal{H}^{\text{eff}}(\mathbf{k})$ are still the eigenstates of rotational operators, which allows us to justify the applicability of Eq.~\ref{eq:SASI} in the amorphous case, by directly observing exchange of high-symmetry eigenvalues between the two lowest-energy bands under application of $\Sigma(\mathbf{k})$. Both calculations are detailed in Appendix \ref{app:Sigma-details}.

\section{Conclusions}
\label{sec:conclusions}

In conclusion, we have shown that for a certain class of amorphous systems, where the local coordination remains constant while  the connectivity can vary by inserting local defects (of Stone-Wales type here), one can infer the topological phase diagram by relying on the reference crystalline topology. For that we introduce the computationally efficient spillage method, requiring only a few eigenstates, that indicates if a band-inversion-like event occurred during amorphization. If it did occur, we can apply a simple crystalline symmetry indicator for the crystal band inversion to successfully predict the topology of the amorphous system.

Finally, we have explained the physical reason behind this ``band inversion'' under amorphization by using a disorder averaged self-energy. In the limit of extremely weak disorder (one defect) we were able to treat the self-energy as perturbation to the crystalline Bloch Hamiltonian, and show that the SW defect effectively introduces sublattice-valley polarization for hopping amplitudes, hence moving the band crossing point from $\pi/3$ to higher values of $\varphi$ and causing the band inversion. 

In the particular case of graphene as a honeycomb lattice, the amorphization by SW defects has been extensively studied (e.g., Ref. [\onlinecite{SWgraphene}]), and it was found that even with $\alpha\approx0.15$ the graph remains hyperuniform \cite{SWuniform}. It is consistent that a method which uses local defects leads to amorphous states that still contain some degree of long-range structure. It would be interesting to consider extended defects and other step-wise routes to generate amorphous states outside the hyperuniform class.

\begin{acknowledgments}
 This work was partially funded by the Region Ile de France through the DIM QUANTIP and by the French Agence Nationale de la Recherche (ANR) under grant number ANR-23-CE30-0037.
 We thank P. d'Ornellas, A. Grushin, Q. Marsal, E. Rio, F. Piechon and T. Desort for enlightening discussions. 
\end{acknowledgments}


\appendix
\begin{widetext}

\section{Analytical results for the crystal: electronic and topological properties}
\label{app:Appendix-analytics}

\begin{figure}[h!]
 \includegraphics[width=0.8\textwidth]{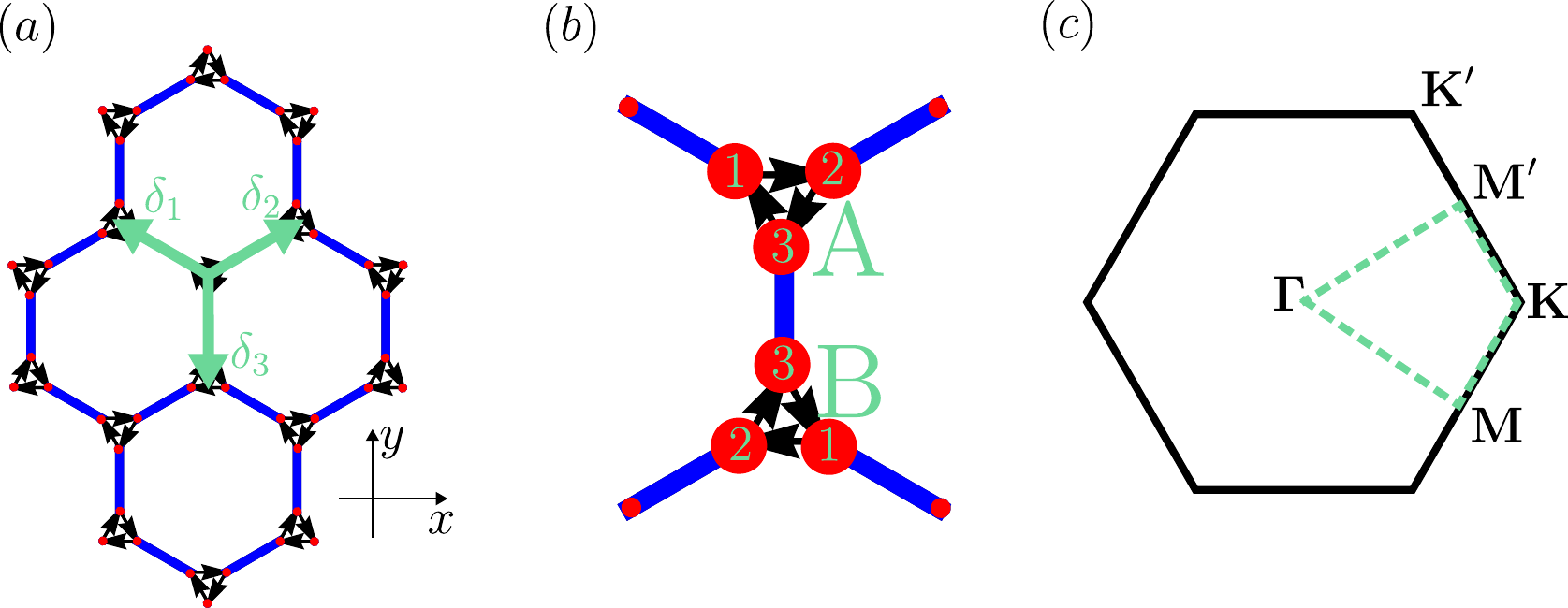}
 \caption{(a) Crystalline lattice with three displacement vectors $\delta_i$. (b) Unit cell with detailed orbital and sublattice labeling. (c) BZ (black) and high-symmetry path (green) used in symmetry indicator calculation.}
\label{fig:crystal_hamiltonian}   
\end{figure}

For the crystalline case, we work in the atomic basis shown in Fig. \ref{fig:crystal_hamiltonian}(a), where we fix the lattice spacing $a=1$, yielding the displacement vectors $\mathbf{\delta}_1 = \left(-\sqrt{3}/2, 1/2\right),~\mathbf{\delta}_2 = \left(\sqrt{3}/2, 1/2\right),~\text{and}~ \mathbf{\delta}_3 = \left(0, -1\right)$. The triangulation of atomic orbitals in Fig.\ref{fig:crystal_hamiltonian}(a,b) is done only for the sake of visual distinguishability. In the Bloch Hamiltonian description we still assume that all three orbitals belong to the same position of the corresponding atom. Therefore, the Bloch Hamiltonian takes the following form:

 \begin{equation}
    \label{eq:Bloch_Hamiltonian} 
{\hat{H}}(\mathbf{k})=
\begin{pmatrix}
0                                   & V e^{i \varphi}                        & V e^{-i \varphi}  & W e^{\frac{-i}{2}(-\sqrt{3} k_x + k_y)} & 0                                       & 0\\
V e^{-i \varphi}                    & 0                                      & V e^{i \varphi}   & 0                                      & W e^{\frac{-i}{2}(\sqrt{3} k_x + k_y)} & 0\\
V e^{i \varphi}                     & V e^{-i \varphi}                       & 0                 & 0                                      & 0                                       & W  e^{i k_y} \\
W e^{\frac{i}{2}(-\sqrt{3} k_x + k_y)}  & 0                                      & 0                 & 0                                      & V e^{i \varphi}                         & V e^{-i \varphi} \\
0                                   & W e^{\frac{i}{2}(\sqrt{3} k_x + k_y)} & 0                 & V e^{-i \varphi}                       & 0                                       & V e^{i \varphi} \\
0                                   & 0                                      & W e^{-i k_y}      &V e^{i \varphi}                         & V e^{-i \varphi}                        & 0\\
\end{pmatrix}
\end{equation}

The symmetry group of this system is the $p6$ wallpaper group, having six-fold rotational symmetry but no real mirror symmetries because of the chirality imposed by the magnetic fluxes $\varphi$. There is however the magnetic mirror, which acts as product of normal mirror and complex conjugation in order to revert the phases in the intra-site hoppings. Rotational symmetry in real space imposes the following commutation relation in momentum space:

\begin{equation}
\label{eq:momentum-pseudo-comutator}
    \hat{R}_n \hat{H}(\mathbf{k})=\hat{H}(R_n \mathbf{k}) \hat{R}_n,
\end{equation}
where $n$ stands for $n$-fold rotation, $\hat{R}_n$ is the corresponding symmetry representation, and $R_n$ is the rotation matrix corresponding to the $n$-fold rotation of the space around $z-$axis:

\begin{equation}
    R_n =   \begin{pmatrix}
\cos (2 \pi / n) &-\sin (2 \pi / n) \\
 \sin (2 \pi / n) &\cos (2 \pi / n)\\
\end{pmatrix} .
\end{equation}

The high symmetry points of the BZ that we label by  the vectors $\mathbf{k}_{\text{inv}}$ satisfy the relation $\hat{H}(R_n \mathbf{k}_{\text{inv}}) = \hat{H}(\mathbf{k}_{\text{inv}} + \mathbf{G})$, where $\mathbf{G}$ is a reciprocal lattice vector. Ideally, one would like to have $\hat{H}(\mathbf{k}_{inv} + \mathbf{G})=\hat{H}(\mathbf{k}_{\text{inv}})$; in this case the relationship in Eq. (\ref{eq:momentum-pseudo-comutator}) would yield a perfect commutator $[\hat{R}_n \hat{H}(\mathbf{k}_{\text{inv}})] =0$ that would significantly simplify the following diagonalization. However, in the \textit{atomic basis} for multi-orbital systems it is not the case, as the Hamiltonian carries inconvenient phases \cite{gillesBasis1Basis2} while moving  from one symmetry point (in our case $\mathbf{K}$) to another equivalent $\mathbf{K}$. For example, in our system written in the atomic basis, the Hamiltonian, at every momentum $\mathbf{k}$, satisfies
\begin{equation}
\label{eq:Hamiltonian-plus-G-relation}
    \hat{H}(\mathbf{k} + \mathbf{G}) = D^*_\mathbf{G} \hat{H}(\mathbf{k}) D_\mathbf{G},
\end{equation}
where $D_\mathbf{G}$ is a diagonal matrix given by
\[
D_\mathbf{G} = \text{diag}(e^{-i \mathbf{G}\mathbf{\delta}_1},~e^{-i \mathbf{G}\mathbf{\delta}_2},~e^{-i \mathbf{G}\mathbf{\delta}_3},~e^{-2i \mathbf{G}\mathbf{\delta}_1},~e^{-2i \mathbf{G}\mathbf{\delta}_2},~e^{-2i \mathbf{G}\mathbf{\delta}_3}).
\]
One can overcome this issue and make the Hamiltonian periodic in $\mathbf{k}$ by going to the cell-centered basis with the momentum-dependent transformation $A(\mathbf{k})$:
\[
A(\mathbf{k})= 
    \begin{pmatrix}
e^{i\frac{k_y}{2}}\mathbb{1}_{3\times3} &0 \\
0 &e^{-i\frac{k_y}{2}}\mathbb{1}_{3\times3}\\
\end{pmatrix} ,
\]
and
\begin{equation}
\label{eq:hamiltonians_basis1_basis2}
    \tilde{H}(\mathbf{k}) = A^\dagger(\mathbf{k}) \hat{H}(\mathbf{k})A(\mathbf{k}),
\end{equation}
where $\tilde{H}$ is the Bloch Hamiltonian in the cell-centered basis which recovers the momentum periodicity: $\tilde{H}(\mathbf{k}_{inv} + \mathbf{G})=\tilde{H}(\mathbf{k}_{\text{inv}})$.

Hereafter, we will mark all the operators that are defined in the atomic basis with a hat: $\hat{\mathcal{O}}$, and operators  in the cell-centered basis with a tilde: $\tilde{\mathcal{O}}$.  Starting from the symmetry commutation condition in 
Eq. (\ref{eq:momentum-pseudo-comutator}) for arbitrary $\mathbf{k}$, one finds, after some algebra, the following relation for symmetry operators between different bases:

\[
    \hat{H}(\mathbf{k}) = A(\mathbf{k}) \tilde{H}(\mathbf{k})A^\dagger(\mathbf{k}),
\]
\[
    \hat{R}_n A(\mathbf{k}) \tilde{H}(\mathbf{k})A^\dagger(\mathbf{k})=A(R_n \mathbf{k}) \tilde{H}(R_n\mathbf{k})A^\dagger(R_n\mathbf{k}) \hat{R}_n,
\]
\[
       A^\dagger(R_n \mathbf{k}) \hat{R}_n A(\mathbf{k}) \tilde{H}(\mathbf{k}) =  \tilde{H}(R_n \mathbf{k})A^\dagger(R_n \mathbf{k}) \hat{R}_n A(\mathbf{k}),
\]
recovering the commutator:
\[
     \tilde{R}_n(\mathbf{k}) \tilde{H}(\mathbf{k}) =  \tilde{H}(R_n \mathbf{k})\tilde{R}_n(\mathbf{k}),
\]
with rotational operator defined in cell-centered basis: 
\begin{equation}
\label{eq:Symmetriesbas1bas2}
\tilde{R}_n(\mathbf{k})= A^{\dagger}(R_n\mathbf{k}) \hat{R}_n A(\mathbf{k}).
\end{equation}

Now let us discuss the symmetry representations of the two-, three-, and six-fold rotational symmetry in the two above bases. We start from atomic basis and by permuting the orbitals correspondingly to the spatial rotation of the lattice. Since the three and six-fold rotations also involve the three-fold permutations of the orbital labels for each atom, we specify the permutation matrix $C_3$:
\begin{equation}
\label{eq:C3-definition}
C_3 =  
    \begin{pmatrix}
0 & 1 & 0 \\
0 & 0 & 1\\
1 & 0 & 0\\
\end{pmatrix}.
\end{equation}

Then we find the following symmetry representations for the two-, three-, and six-fold rotations, respectively:

\begin{equation}
\hat{R}_2=\begin{pmatrix}
0 &\mathbb{1}_{3\times3} \\
\mathbb{1}_{3\times3} &0\\
\end{pmatrix}  \quad
\hat{R}_3=\begin{pmatrix}
C_3 &0 \\
0 &C_3\\
\end{pmatrix}, \quad
\hat{R}_6=\begin{pmatrix}
0 &C^{-1}_3 \\
C^{-1}_3 &0\\
\end{pmatrix}
\end{equation}

Following Ref. \cite{PointGroupInv}, we chose the same high-symmetry path (Fig. \ref{fig:crystal_hamiltonian}(c))  for the symmetry indicator in Eq. (\ref{eq:symm-ind}). Therefore, the high-symmetry points that will be further  discussed are $\mathbf{\Gamma}=\{0,0\}$, $\mathbf{K} = \{\frac{4\pi}{3\sqrt{3}},0\}$, and $\mathbf{M} = \{{\frac{\pi}{\sqrt{3}}, -\frac{\pi}{3}}\}$.  Then, in the cell-centered basis, according to Eq. (\ref{eq:Symmetriesbas1bas2}), the symmetry representations take the following form:

\begin{equation}
\tilde{R}_2(\mathbf{M})=\begin{pmatrix}
0 &\mathbb{1}_{3\times3} \\
\mathbb{1}_{3\times3} &0\\
\end{pmatrix}  \quad
\tilde{R}_3(\mathbf{K})=\begin{pmatrix}
e^{-i\frac{\pi}{3}}C_3 &0 \\
0 &e^{i\frac{\pi}{3}}C_3\\
\end{pmatrix} \quad
\tilde{R}_6 (\mathbf{\Gamma})=\begin{pmatrix}
0 &C^{-1}_3 \\
C^{-1}_3 &0\\
\end{pmatrix} .
\end{equation}

\begin{figure}[h!]
 \includegraphics[width=1\textwidth]{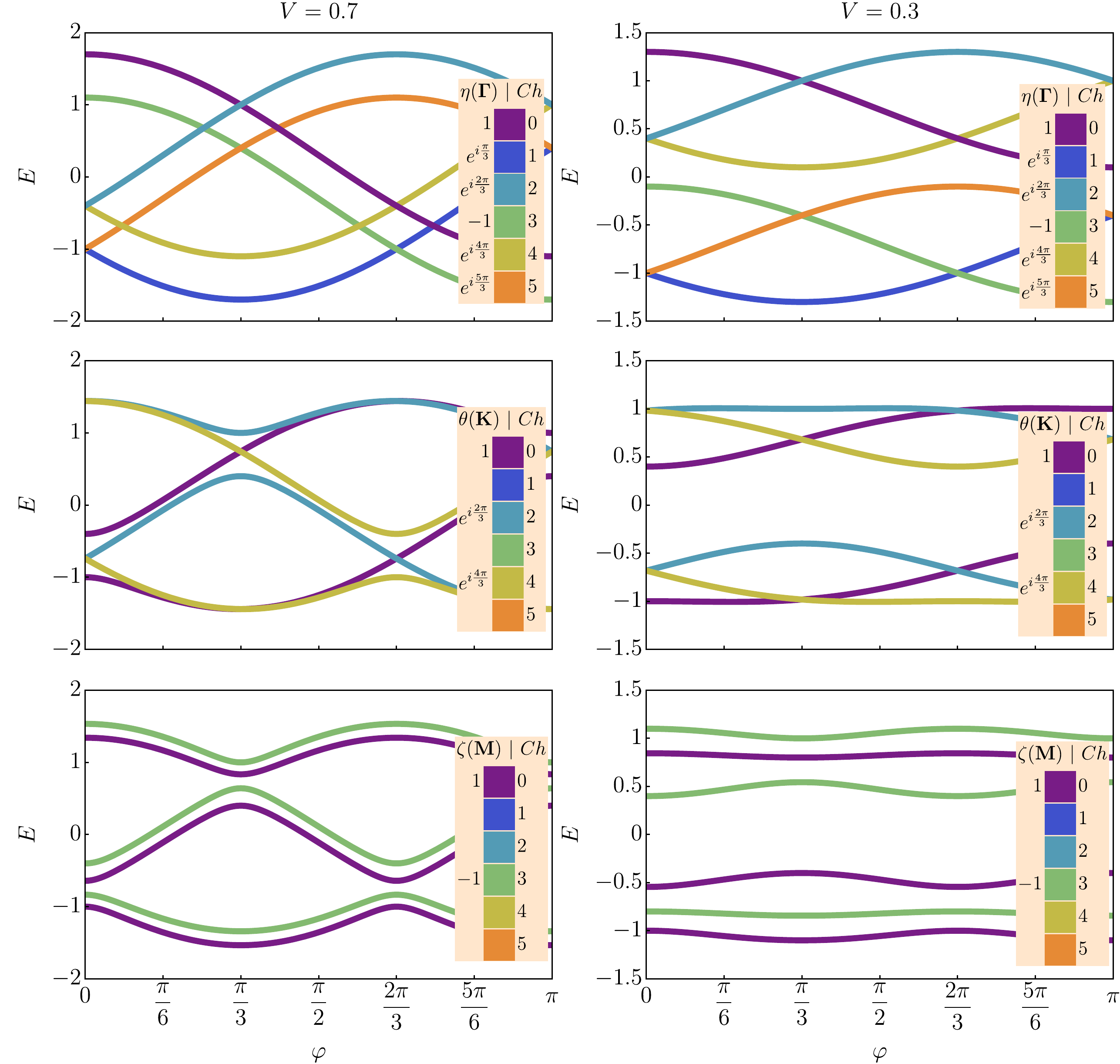}
 \caption{Energy levels dependence on $\varphi$ (given by Eqs. (\ref{eq:AnaEigenEnergies}))  evaluated at the high-symmetry points $\mathbf{\Gamma},~ \mathbf{K},~ \text{and} ~\mathbf{M}$ are shown  from the bottom to top row respectively. Two characteristic regimes $V=0.7$, and $V=0.3$ are depicted in the left and right columns respectively. Colors represent the Chern number contribution carried by Bloch eigenstates at each high symmetry momenta $\mathbf{k_\text{inv}}$ given by Eq. (\ref{eq:ChernContributions}). Chern number contributions (from 0 to 5) are indicated in the right side of the inset legend. Corresponding symmetry eigenvalues are indicated in the left side of the legend. Then, one recovers the whole Chern number (mod 6) by counting the Chern number contributions for all occupied bands and high-symmetry momenta by using Eq. (\ref{eq:chern-as-sum}).}
\label{fig:analytical_solutions}   
\end{figure}

We notice that all symmetry representations are composed of block structures relying on either the $C_3$ operator acting inside the space of the three orbitals for each atom, and/or $\tau_x\equiv\hat{R}_2$ which is a Pauli matrix acting as an exchange between A and B "sublattices". We therefore  label the states of the  Bloch Hamiltonian with indices $m$, and $\nu$ which correspond to different eigenvalues for $C_3$ and $\tau_x$, respectively. More specifically,  the eigenvalues of the $C_3$ matrix defined in Eq. (\ref{eq:C3-definition}) are given by $\lambda_m = e^{i\frac{2\pi}{3}m}$, with corresponding eigenvectors 
\begin{equation}
v_m = \frac{1}{\sqrt3} \left(e^{i\frac{2\pi}{3}m}~,~e^{-i\frac{2\pi}{3}m}~,~1\right)^T,
\end{equation}
with  $m=\pm1,~0$. Concerning the $\tau_x$ operator we introduce the $\pm$ labeling for its respective $\pm1$ eigenvalues.

Having recovered the proper commutation relation $[\tilde{R}_n(\mathbf{k}_\text{inv}), \tilde{H}(\mathbf{k}_\text{inv})] = 0$ at the high symmetry points, we then diagonalize the Hamiltonian in the cell centered basis with the use of a common set of eigenstates between the symmetry representations and the Hamiltonian. At high-symmetry momenta, the eigenenergies take the form:

\begin{equation}
\label{eq:AnaEigenEnergies}
    \begin{cases}
      \mathcal{E}^{(m)}_{\pm}(\mathbf{\Gamma},V,\varphi) = 2V\cos{\left(\varphi-m\frac{2\pi}{3}\right)} \pm W & \quad  \quad m = \pm1,0\\
      
       \mathcal{E}^{(m)}_{\pm}(\mathbf{K},V,\varphi) = V\cos{\left(\varphi+m\frac{2\pi}{3}-\frac{\pi}{3} \right)}\pm\sqrt{3V^2\sin^2{\left(\varphi+m\frac{2\pi}{3}-\frac{\pi}{3}\right)}+W^2}& \quad  \quad m = \pm1,0\\
       \mathcal{E}^{(m)}_{\pm}(\mathbf{M},V,\varphi) = \frac{2}{3}\sqrt{p}\cos{\left[\frac{1}{3}
    \arccos\left(\frac{q_{\mp}(\varphi)}{p^{3/2}}\right)+m\frac{2\pi}{3}\right]}\pm \frac{W}{3} & \quad   \quad m = \pm1,0,\\
    \end{cases}
\end{equation}
where $p = (3V)^2+(2W)^2$, and $q_{\mp}(\varphi)=(3V)^3\cos{(3\varphi)\mp(2W)^3}$. 
They are represented in Fig. \ref{fig:analytical_solutions}
with their corresponding eigenstates $\tilde{\psi}$ that read in the cell centered basis as
\begin{equation}
\label{equation:gammaeigenstates}
    \mathcal{E}^{(m)}_{\pm}(\mathbf{\Gamma},V,\varphi) 	\Leftrightarrow \ket{\tilde{\psi}^{(m)}_{\pm}(\mathbf{\Gamma},V,\varphi)}=\frac{1}{\sqrt{2}}\left(v_m~,~\pm v_m\right)^T,
\end{equation}
\begin{equation}
\label{equation:keigenstates}
\mathcal{E}^{(m)}_{\pm}(\mathbf{K},V,\varphi) 	\Leftrightarrow \ket{\tilde{\psi}^{(m)}_{\pm}(\mathbf{K},V,\varphi)}=\frac{1}{\sqrt{1+|\alpha^{\pm}_{m}|^2}}\left(\alpha^{\pm}_{m}v_{-m}~,~ v_{-(m-1)}\right)^T,
\end{equation}
where indexing $m-1$ should be understood  modulo 3 with an offset -1 (i.e. $m=-2 \Leftrightarrow m=1; ~m=2 \Leftrightarrow m=-1$), and $ \alpha^{\pm}_{m} =  (\mathcal{E}^{(m)}_{\pm}(\mathbf{K},V,\varphi)-2V \cos{(\varphi + \frac{2\pi}{3}(m-1))})/W$.
Finally, the eigenstate at the $\mathbf{M}$ point reads 
\begin{equation}
\label{equation:meigenstates}
\mathcal{E}^{(m)}_{\pm}(\mathbf{M},V,\varphi) 	\Leftrightarrow \ket{\tilde{\psi}^{(m)}_{\pm}(\mathbf{M},V,\varphi)}=\frac{1}{\sqrt{2|\omega_m|^2}}\left(\gamma^{\pm}_m~,~\beta^{\pm}_m~,~1~,
~\pm\gamma^{\pm}_m~,~\pm\beta^{\pm}_m~,~\pm 1\right)^T,
\end{equation}
where $\gamma^{\pm}_m =e^{2i\varphi} (Ve^{-3i\varphi} \mp W + \mathcal{E}^{(m)}_{\pm}(\mathbf{M},V,\varphi))/(Ve^{3i\varphi} \pm W + \mathcal{E}^{(m)}_{\pm}(\mathbf{M},V,\varphi))$, and $\beta^{\pm}_m = \frac{e^{i\varphi}}{V}(-Ve^{i\varphi}\gamma^{\pm}_m\mp W+\mathcal{E}^{(m)}_{\pm}(\mathbf{M},V,\varphi))$.

Then, by applying Eq. (\ref{eq:hamiltonians_basis1_basis2}) we recover the eigenstates $\psi(\mathbf{k})$ in the atomic basis:

\begin{equation}
\ket{\psi^{(m)}_{\pm}(\mathbf{\Gamma},V,\varphi)}\equiv\ket{\tilde{\psi}^{(m)}_{\pm}(\mathbf{\Gamma},V,\varphi)},
\end{equation}

\begin{equation}
\ket{\psi^{(m)}_{\pm}(\mathbf{K},V,\varphi)} = A(\mathbf{K})\ket{\tilde{\psi}^{(m)}_{\pm}(\mathbf{K},V,\varphi)} = \ket{\tilde{\psi}^{(m)}_{\pm}(\mathbf{K},V,\varphi)},
\end{equation}

\begin{equation}
\ket{\psi^{(m)}_{\pm}(\mathbf{M},V,\varphi)} = A(\mathbf{M})\ket{\tilde{\psi}^{(m)}_{\pm}(\mathbf{M},V,\varphi)} = \begin{pmatrix}
e^{-i\frac{\pi}{6}}\mathbb{1}_{3\times3} &0 \\
0 &e^{i\frac{\pi}{6}}\mathbb{1}_{3\times3}\\
\end{pmatrix} \ket{\tilde{\psi}^{(m)}_{\pm}(\mathbf{M},V,\varphi)}. 
\end{equation}

Using Eq. (\ref{eq:Hamiltonian-plus-G-relation}), we also recover momentum-rotated 
eigenstates $\psi(R_n \mathbf{k}_{\text{inv}})$:

\begin{equation}
\ket{\psi^{(m)}_{\pm}(R_n \mathbf{k_{\text{inv}}},V,\varphi)} =
D^*_\mathbf{G}\ket{\psi^{(m)}_{\pm}( \mathbf{k_{\text{inv}}},V,\varphi)}.
\end{equation}

We remind that for calculating the symmetry indicator in Eq. (\ref{eq:symm-ind}) one needs the eigenvalues of rotational operators at high-symmetry points: 

\begin{equation}
Ch_{\text{ind}} =\frac{3}{i\pi}\log\left[\prod_{i \in o c c .} \eta_i(\mathbf{\Gamma}) \theta_i(\mathbf{K}) \zeta_i(\mathbf{M})\right] ~\text{mod} ~6.
\end{equation}

To find the rotational eigenvalues we apply the sewing matrix formalism \cite{PointGroupInv}, which is a diagonal matrix at high-symmetry momenta:

\begin{equation}
\label{eq:sewing_matrix_def}
\left(\mathcal{B}_{R_n}(\mathbf{k})\right)_{\mu \nu}=\bra{\psi_{\mu}(R_n \mathbf{k})} \hat{R}_n\ket{\psi_{\nu}(\mathbf{k})},
\end{equation}
where $\mu$ and $\nu$ are generic band indices, that include $\pm,~\text{and}~ m$ labeling in our case. At high-symmetry momenta, the sewing matrix is exactly diagonal, with entries that correspond to rotational operators' eigenvalues:

\begin{equation}
\left(\mathcal{B}_{R_n}(\mathbf{k}_{\text{inv}})\right)_{\mu \nu} = \Lambda^{\mu}_{R_{n}}(\mathbf{k}_{\text{inv}}) \delta_{\mu \nu},
\end{equation}
where $\Lambda^{\mu}_{R_{n}}(\mathbf{k}_{\text{inv}})$ is the eigenvalue of $\hat{R}_{n}$ evaluated for band $\mu$ at the high-symmetry momentum $\mathbf{k}_{\text{inv}}$.

By computing the sewing matrices (Eq. \ref{eq:sewing_matrix_def}) for  $\mathbf{\Gamma},~\mathbf{K},~\text{and}~\mathbf{M}$, we find the following rotational eigenvalues for each value of $m$, and $\pm$:
\[
    \eta^{(m)}_{\pm}(\mathbf{\Gamma}) = \bra{\tilde{\psi}^{(m)}_{\pm}(\mathbf{\Gamma},V,\varphi)} \hat{R}_6 \ket{\tilde{\psi}^{(m)}_{\pm}(\mathbf{\Gamma},V,\varphi)},
\]
\[
    \theta^{(m)}_{\pm}(\mathbf{K}) = \bra{\tilde{\psi}^{(m)}_{\pm}(\mathbf{\mathbf{K}},V,\varphi)} D_\mathbf{G} \hat{R}_3 \ket{\tilde{\psi}^{(m)}_{\pm}(\mathbf{\mathbf{K}},V,\varphi)},
\]
\[
     \zeta^{(m)}_{\pm}(\mathbf{M}) = \bra{\tilde{\psi}^{(m)}_{\pm}(\mathbf{\mathbf{M}},V,\varphi)} A(\mathbf{M}) D_\mathbf{G} \hat{R}_2 A^\dagger(\mathbf{M}) \ket{\tilde{\psi}^{(m)}_{\pm}(\mathbf{\mathbf{M}},V,\varphi)},
\]
where in both cases for selected $\mathbf{M}$, and $\mathbf{K}$, the $\mathbf{G}=\{-\frac{2\pi}{\sqrt{3}}, \frac{2\pi}{3}\}$. After some algebra, the eigenvalues are then simplified to the following:

\begin{equation}
\label{eq:SymmEigenval}
    \begin{cases}
    \eta^{(m)}_{\pm}(\mathbf{\Gamma})  =  \pm e^{i\frac{2\pi}{3}m}\\
    \theta^{(m)}_{\pm}(\mathbf{K}) = e^{i\frac{2\pi}{3}(m+1)}\\
   \zeta^{(m)}_{\pm}(\mathbf{M}) = \mp 1\\
    \end{cases}~~~.
\end{equation}
This implies that these eigenvalues are $V\text{-},~\text{and}~\varphi$-independent in general, and in particular, $\theta^{(m)}_{\pm}(\mathbf{K}) $ are $\pm$-independent, and $ \zeta^{(m)}_{\pm}(\mathbf{M})$ are $m$-independent. Having the eigenvalues, one thus can compute the contribution to the Chern number for each band $\mu$ at each high-symmetry momenta by formula:

\begin{equation}
\label{eq:ChernContributions}
Ch_{\mu}(\mathbf{k_\text{inv}}) = \frac{3}{i\pi}\log \Lambda^{\mu}_{R_n}(\mathbf{k_\text{inv}})~\text{mod} ~6,    
\end{equation}

The resulting symmetry eigenvalues, and Chern number contribution are presented in Fig. \ref{fig:analytical_solutions} for each high-symmetry momenta. This makes the computation of Chern number straightforward: 

 \begin{equation}
\label{eq:chern-as-sum}
Ch_{\text{ind}} =\sum_{\mu \in o c c .}\sum_{\mathbf{k_\text{inv}}} Ch_{\mu}(\mathbf{k_\text{inv}}) ~\text{mod} ~6.
\end{equation}.

For example, for the filling factor $f=1/6$ discussed in this work and chosen range of $0<\varphi<\pi/3$, one has to perform the summation only over the first band:

\[
Ch_{\text{ind}} = Ch_{1}(\mathbf{\Gamma})+Ch_{1}(\mathbf{K})+Ch_{1}(\mathbf{M}) ~~\text{mod} ~6 = 1+0+0 ~~\text{mod} ~6 = 1 ~~\text{mod} ~6.
\]

By repeating the procedure for different values of $\varphi$ and $V$, one recovers the topological phase diagram presented in Fig.\ref{fig:main-result}(a).

\section{Lattice amorphization}
\label{app:Appendix-lattice-amorphisation}
\begin{figure}[h!]
 \includegraphics[width=1\textwidth]{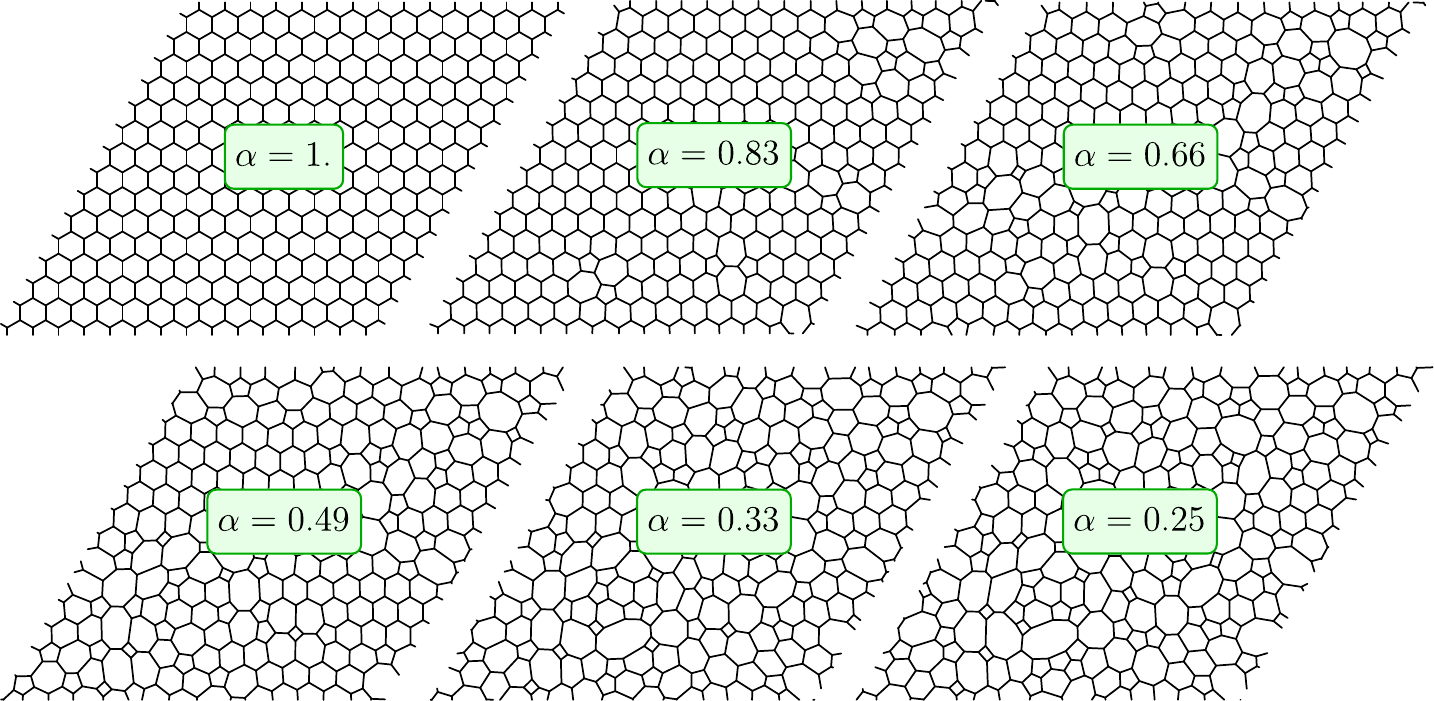}
 \caption{Typical amorphisation evolution from crystalline $(\alpha=1)$ to amorphous lattice $(\alpha=0.25)$ for system of size $15\times15$ u.c. simulated with Surface Evolver program with lattice relaxation constraints given by \ref{eq:target-surface-areas}.}
\label{fig:lattice_example}   
\end{figure}

To produce a continuous deformation process that would drive the system from the crystalline to the amorphous state, we use a consecutive application of the SW defect (Fig. \ref{fig:models}) to the different randomly chosen inter-atomic bonds.

We use the Surface Evolver program introduced in \cite{brakke1992surface} to programmatically realize the amorphization procedure. This program allows to efficiently mesh and model surfaces in different dimensions. In our case, we use it to create a periodic hexagonal mesh of the torus (Fig. \ref{fig:lattice_example} ), amorphize it with use of SW procedure described above, and then relax a mesh to the visually appealing state. Surface Evolver allowed us to keep the planarity of the graph embedding of the torus, while maintaining reasonable proportionality of the plaquettes' surface areas to number of the bonds constructing the loop (i.e. the valence of plaquette), given by the formulae:

\begin{equation}
\label{eq:target-surface-areas}
    \begin{cases}
      S \propto (\frac{L}{6})^4&\text{for}\quad L \leq 5\\
      S \propto(\frac{L}{6})^2&\text{for}\quad L \geq 6,\\
    \end{cases}
\end{equation}
where $S$ and $L$ are the target surface area, and the valence of the plaquette, respectively. With these conditions, we fix the characteristic surface area to be 1 when plaquette is hexagonal, and we penalize the small-valence plaquettes in their surface more strongly to create more visually appealing graph. Then we normalize the target surface areas of the plaquettes in order to keep the total surface area of the torus to be 1: $\sum_{i \in \text{plaq.}}S_i=1$. After fixing the target surfaces of the plaquettes, Surface Evolver then relaxes the graph in order to converge current surface areas to the target surface areas, and to minimize the spring potential energy assigned to the edges of the graph. Graph relaxation and readout of the resulting positions of the atoms is done on each step of amorphisation. Importantly, the atomic labeling is kept intact throughout the whole process of amorphisation, making computation of overlaps of eigenstates for different values of $\alpha$ possible and meaningful, which is used for evaluation of the  eigenstate spillage  defined in Eq. (\ref{equation:MySpillage-full}).

\section{Range of applicability of eigenstate spillage indicator, and statistical deviations due to disorder averaging, and ambiguity of bond rotation direction}
\label{app:Appendix-spillage-errors}

In this section, we discuss the range of topological disorder, where the eigenstate spillage indicator remains meaningful. To do so, we evaluate the indicator for the whole range of parameters $W\in[0,1]$, and $\varphi\in[0,\pi]$, and for a set of disorder parameters alpha: $\alpha = 3\%, ~18\%, ~33\%, ~45\%, \text{and} ~60\%$. The disorder-averaged (50 disorder realizations), for a lattice size 12$\times$12 u.c., are presented in Fig. \ref{fig:spillage_phase_diags_as_func_of_disorder}:

\begin{figure}[h!]
 \includegraphics[width=1\textwidth]{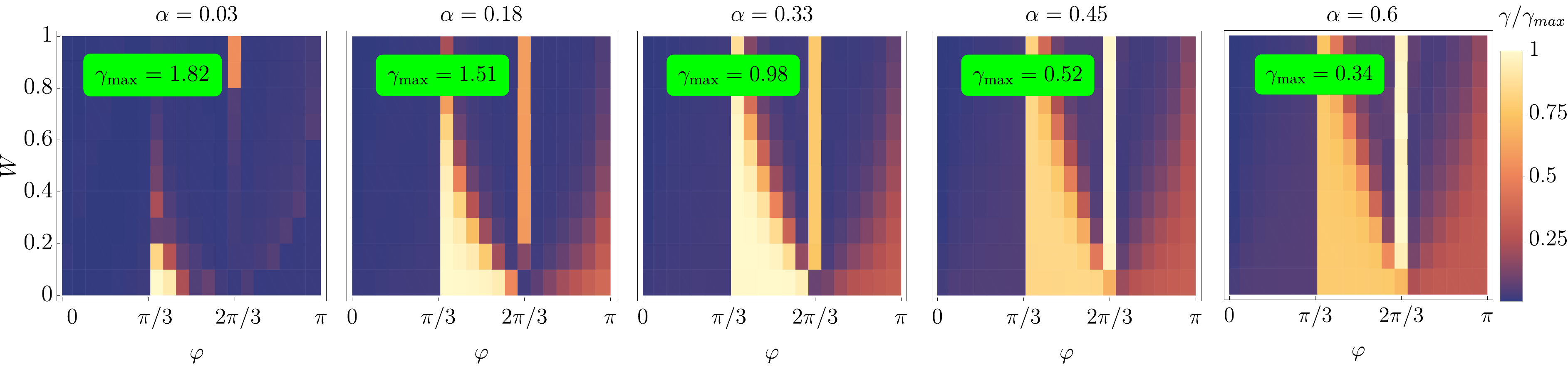}
 \caption{Eigenstate spillage phase diagrams as function of disorder strength $\alpha$ from left to right: $\alpha = 3\%, ~18\%, ~33\%, ~45\%, \text{and} ~60\%$. Each diagram is normalized to the maximum value of the spillage $\gamma_{\text{max}}$ in order to make meaningful comparison between different disorder strengths. For each diagram $\gamma_{\text{max}}$ is indicated in a green box inside the plot.}
\label{fig:spillage_phase_diags_as_func_of_disorder}   
\end{figure}

Each diagram is normalized to the maximum value of spillage (green box inset) for each disorder strength, to keep  the same color scale for all disorder regions. One can observe that going from weak disorder ($\alpha = 3\%$) to strong disorder ($\alpha = 60\%$) the maximum value of the spillage almost decreases by one order of magnitude due to an increase of the distance in Hilbert space between amorphous and crystalline eigenstates with  disorder. Because of that, no universal constant cutoff for different disorder strength can be introduced to distinguish between the phases. Regardless of that, the different regions of the phase diagrams remain quite distinguishable up to disorder values of $\alpha = 30\% \sim 40\%$.

In order to determine more precisely the disorder strength where the regions of different phases can still be distinguished, we  analyze the  error accumulation due to disorder averaging. We fix $W=0.2$ in the rest of this section and compare the system  in three different regimes parametrized by $\varphi=\pi/3+0.1$ (deep interior of the phase region of pseudo-band inversion), $\varphi=\pi - 0.1$  (deep interior point of the phase region of disorder-localized phase), and $\varphi=\pi/3 - 0.1$  (deep interior point of the phase region where no phase transition occurs). In the crystalline limit, these three phases have the same energy spectra. We estimate the typical error due to disorder averaging by computing standard deviation:
\begin{equation}
    \label{eq:standard_deviation}
    \sigma_{\gamma}(\alpha) = \sqrt{\frac{1}{N-1} \sum^{N}_{i=1}(\gamma_i - \bar{\gamma})^2},
\end{equation}
\begin{figure}[h!]
\includegraphics[width=1\textwidth]{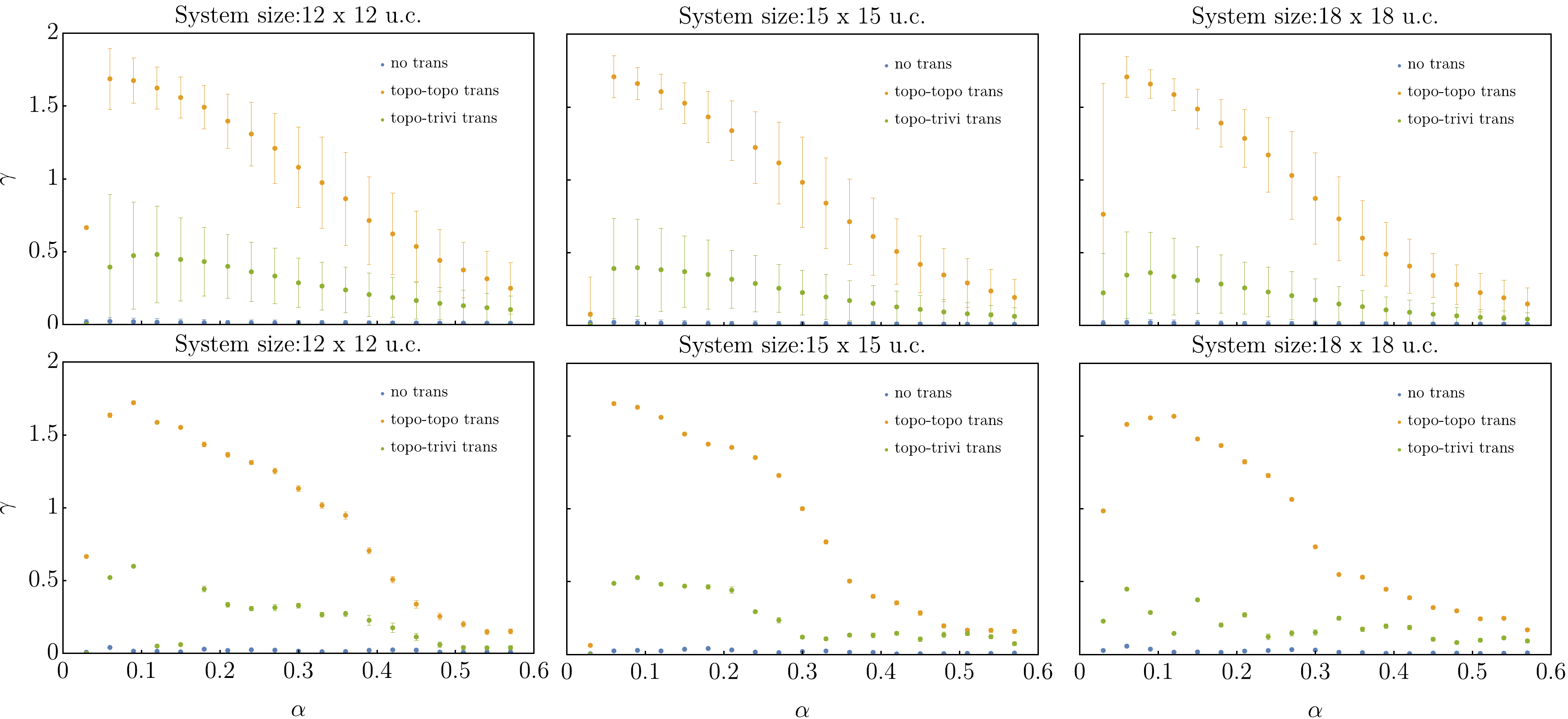}
\caption{Eigenstate spillage evolution with varied disorder strength $\alpha=0.03..0.6$. Panels from left to right for 12$\times$12, 15$\times$15, 18$\times$18 u.c. system sizes. Dots indicate the disorder-averaged mean value, and error bars indicate the confidence interval of 95\% - two standard deviations computed by \ref{eq:standard_deviation}. Upper panels (300 configurations): mean and two standard deviations due to disorder averaging. Lower panels (100 samples, same disorder configuration): mean and two standard deviations for one fixed disorder configuration but randomized directions of $W-$bond rotation under SW process.}
\label{fig:errorBars_disorder}
\end{figure}
where $N$ is the number of disorder configurations and $\bar{\gamma}$ is an average of $\gamma_{i}$ for fixed $\alpha$. We also perform a scaling analysis for system sizes of: 12$\times$12, 15$\times$15, and 18$\times$18 u.c.. We keep the number of disorder configurations fixed  to 300 between different system sizes to avoid artificial underestimation of the error due to reduction of the number of samples for larger systems. The results are presented in the upper panel of Fig.\ref{fig:errorBars_disorder}. We indicate the mean value of spillage as function of $\alpha$ by orange, green and blue dots of different colors, for the corresponding three scenarios: transition from the topological phase $\text{Chern} = -1$ to the topological phase: $\text{Chern} = 1$ ($\varphi=\pi/3+0.1$), transition from the topological phase $\text{Chern} = 1$ to a trivial phase $\text{Chern} = 0$ ($\varphi=\pi - 0.1$), no phase transition at all ($\varphi=\pi/3 - 0.1$). The corresponding error bars indicate the confidence interval of 95\% corresponding to two standard deviations computed by \ref{eq:standard_deviation}.
As one can see, the discrimination between the regime of topological-topological transitions (orange dots) and other regimes remains reliable for up to $\alpha\approx0.35$. For stronger disorder one has some finite probability to misclassify the phase as being topological-trivial one (green dots). Scaling analysis (going in panels from left to right) demonstrates a slow decrease of standard deviations with the increase of system size, and correspondingly a better discrimination between topological-topological transition and other transitions.

Concerning the  topological-trivial transition and no transition, even though the no-transition case (blue dots) has tiny error bars, the discrimination between the regimes may become problematic for any disorder strength if it statistically happens that spillage for topological-trivial case has a value close to zero. In this case, the disorder averaging is necessary for being able to discriminate between the two phases. Therefore, in the main text (Fig. \ref{fig:main-result}(b)) we only refer to disorder-averaged value of spillage.

We also assessed the error accumulation due to the ambiguity in the direction of $W-$bond rotation of the SW defect. Since all atoms in our model have identical local environment, a clockwise and anti-clockwise rotation of $W-$bond produces two isomorphic graphs, though the two atoms exchange their labels. This means that by randomizing the $W-$bond directions of rotation we produce a set of Hamiltonians, that have the same spectra, but their eigenstates have swapped labels for the orbitals affected by the SW process. Therefore, even if disorder configuration stays the same, one can expect different eigenstate spillage value for systems with randomized SW defect directions. Indeed, for example, for only two defects in the system (say affecting pairs of atoms $A-B$, and $C-D$), one can see the differences in computation of the overlaps between amorphous and crystalline eigenstates. If the original spillage would depend on the overlap as
\[
\gamma_{\text{orig}}\propto|\bra{\psi_{\text{cry}}}\ket{\psi_{\text{amo}}}|^2,
\]
then the system with another bond rotation direction would have the overlap
\[
\gamma_{\text{rot}}\propto|\bra{\psi_{\text{cry}}}\mathbf{P}_{\text{AB}}\mathbf{P}_{\text{CD}}\ket{\psi_{\text{amo}}}|^2,
\]
where $\mathbf{P}$ are the permutation matrices between labels of two SW-affected atoms. Clearly, $\gamma_{\text{orig}} \neq \gamma_{\text{rot}}$, therefore some error is accumulated because of this ambiguity. To assess it, we again refer to three different system sizes of: 12$\times$12, 15$\times$15, 18$\times$18 u.c., we fix one disorder configuration, and randomize the directions of $W-$bond rotations at each step of amorphisation. With that, we generate 100 samples and compute the error by using standard deviation \ref{eq:standard_deviation}. The results are presented in lower panels of Fig. \ref{fig:errorBars_disorder}, where error bars again indicate the interval of two standard deviations. As one can see, the errors due two SW rotation ambiguity (lower panels) are \textit{negligible} w.r.t. the errors generated by sampling the systems with different disorder configurations (upper panels). Furthermore, no significant growth of SW-rotation error is observed with increase of $\alpha$.

\section{Details of self-energy computations and physical implications}
\label{app:Sigma-details}

In this section, we discuss the details of computation and further physical implications of the effective self-energy introduced in Sec. \ref{sec:sigma}. In particular, we discuss the recovery of original crystalline symmetries for self-energy, as well as the application of self-energy to Bloch Hamiltonian which causes an inversion of the Berry curvature, and a corresponding inversion of high-symmetry eigenvalues.

In order to generate our configurations, over which we perform the configurational averaging of T-matrix, we do the following: we take a crystalline system and introduce only one SW defect to some $W-$bond, for which we compute a local defect potential and evaluate a corresponding T-matrix by eq. \ref{eq:T-matrix}, then we take another bare crystalline system, and introduce again only one defect but this time SW procedure is applied to another $W-$bond. We repeat this procedure for all $W-$bonds present in the crystal. This generates us $3\times N_{u.c.}$ configurations of crystalline lattices, each with one defect applied to a certain $W-$bond (as there are three $W-$bonds per unit cell). Then we repeat this procedure, but now we rotate our defect in opposite direction, since an SW defect has built-in ambiguity in bond rotation direction, which doubles our number of configurations. Therefore, we get $3\times2\times N_{u.c.}$ disorder configurations, and for each configuration we compute a T-matrix. Then we perform an arithmetical averaging and obtain $\langle T \rangle_c$, and consequently $\Sigma$ by eq. \ref{eq:self-ene}. For computation of the Green's function of a crystal $G_0$, we set the Fermi energy $\varepsilon_F$ to be in the middle of the mini-gap of interest, corresponding to the filling factor $f=1/6$, and fix $\eta=0.003$. After finding the self-energy we take the Hermitian part of is $(\Sigma+\Sigma^{\dagger})/2$, as we are interested only in coherent renormalization of the bands, and perform a Fourier transform by projecting onto plane-wave basis. In our case the self-energy is exactly diagonal in the momentum space, which allows us to treat it as a small perturbation to the Bloch Hamiltonian directly in momentum space and analyze how it affects the electronic and geometric properties of the bands.

For the rest of discussion in this Appendix, we also fix the parameters $V=0.8$, and $\varphi=\pi/3+0.05$ to put the system close to the crystalline topological transition point $\varphi_c=\pi/3$. After finding the T-matrix for each configuration, we perform the disorder averaging and find the corresponding self-energy by using eq. \ref{eq:self-ene}. 

This approach is limited to use only for smaller system sizes, as, by going to the thermodynamic limit, the effect of a single defect on bulk properties will be diminished, and one needs to consider some finite concentration of defects. With this in mind, we need to choose the system sizes small enough for allowing the Hamiltonian to experience the effect of self-energy, but large enough for numerical stability and ability to make fine meshes in BZ, as we are interested in  computation of gradients in momentum space. For that purpose, we fix the system size to be $18\times18$ u.c., resulting in 1944 disorder configurations. For simplicity, we also keep the original crystalline positions for the atomic orbitals and do not move two SW-affected atoms but only reconnect $W-$bonds corresponding to SW defect, making the effect of disorder only appear inside the $\mathcal{H}_{\text{def}}$ matrices in eq. \ref{eq:T-matrix}.

\subsection{Recovery of original crystalline point-group symmetries, and components of self-energy over entire BZ}

Our method of averaging over the system's configurations allows us to recover not only the translational invariance of the self-energy, but also all three of the system's rotational symmetries, $\hat{R}_n$, together with magnetic mirrors $\hat{M}_x,$ and $\hat{M}_y$. 
These symmetries are achieved by placing the mirroring centre at the midpoint of the $W-$bond, and positioning two mirrors, $\hat{M}_x$ and $\hat{M}_y$, parallel to and perpendicular to the direction of the $W-$bond.  After mirroring, the time-reversal operator is applied (in our case it is a simple complex conjugation, $\mathcal{K}$). To numerically verify that all mentioned symmetries are restored, we compute the Frobenius norms $||A||_{F} = \sqrt{\operatorname{Tr}(AA^{\dagger})}$ of the commutation relations, which indeed are equal to zero for \textit{all} allowed momenta in the BZ,

\begin{equation}
    ||\cdot||_n=||\hat{R}_n \Sigma(\mathbf{k})-\Sigma(R_n \mathbf{k}) \hat{R}_n||_{F}=0 \quad\quad \forall\mathbf{k}\in\text{BZ},~ \forall n\in[2,3,6],
\end{equation}

\begin{equation}
\label{eq:Mx}
    ||\cdot||_x=||\hat{M}_x \Sigma^*(-k_x, -k_y) - \Sigma(k_x, -k_y) \hat{M}_x||_{F}=0\quad\quad \forall\mathbf{k}\in\text{BZ}~,
\end{equation}

\begin{equation}
\label{eq:My}
    ||\cdot||_y=||\hat{M}_y \Sigma^*(-k_x, -k_y) - \Sigma(-k_x, k_y) \hat{M}_y||_{F}=0\quad\quad \forall\mathbf{k}\in\text{BZ}~,
\end{equation}
where the operators, $\hat{M}_x$ and $\hat{M}_y$, have the following representation:
\begin{equation}
\hat{M}_x=\begin{pmatrix}
0 &\mathbf{M} \\
\mathbf{M} &0\\
\end{pmatrix} ~, \quad
\hat{M}_y=\begin{pmatrix}
\mathbf{M} &0 \\
0 &\mathbf{M}\\
\end{pmatrix}
\end{equation}
with $\mathbf{M}=  
    \begin{pmatrix}
0 & 1 & 0 \\
1 & 0 & 0\\
0 & 0 & 1\\
\end{pmatrix}$ that exchanges orbitals 1 and 2. The recovery of all crystalline symmetries after configurational averaging makes $\Sigma(\mathbf{k})$ a simple perturbation over $\mathcal{H}_{\text{cry}}$,  allowing us to study the geometrical properties of the resulting bands using symmetry indicators (see the second subsection of this Appendix).

\begin{figure}[h!]
\includegraphics[width=0.75\textwidth]{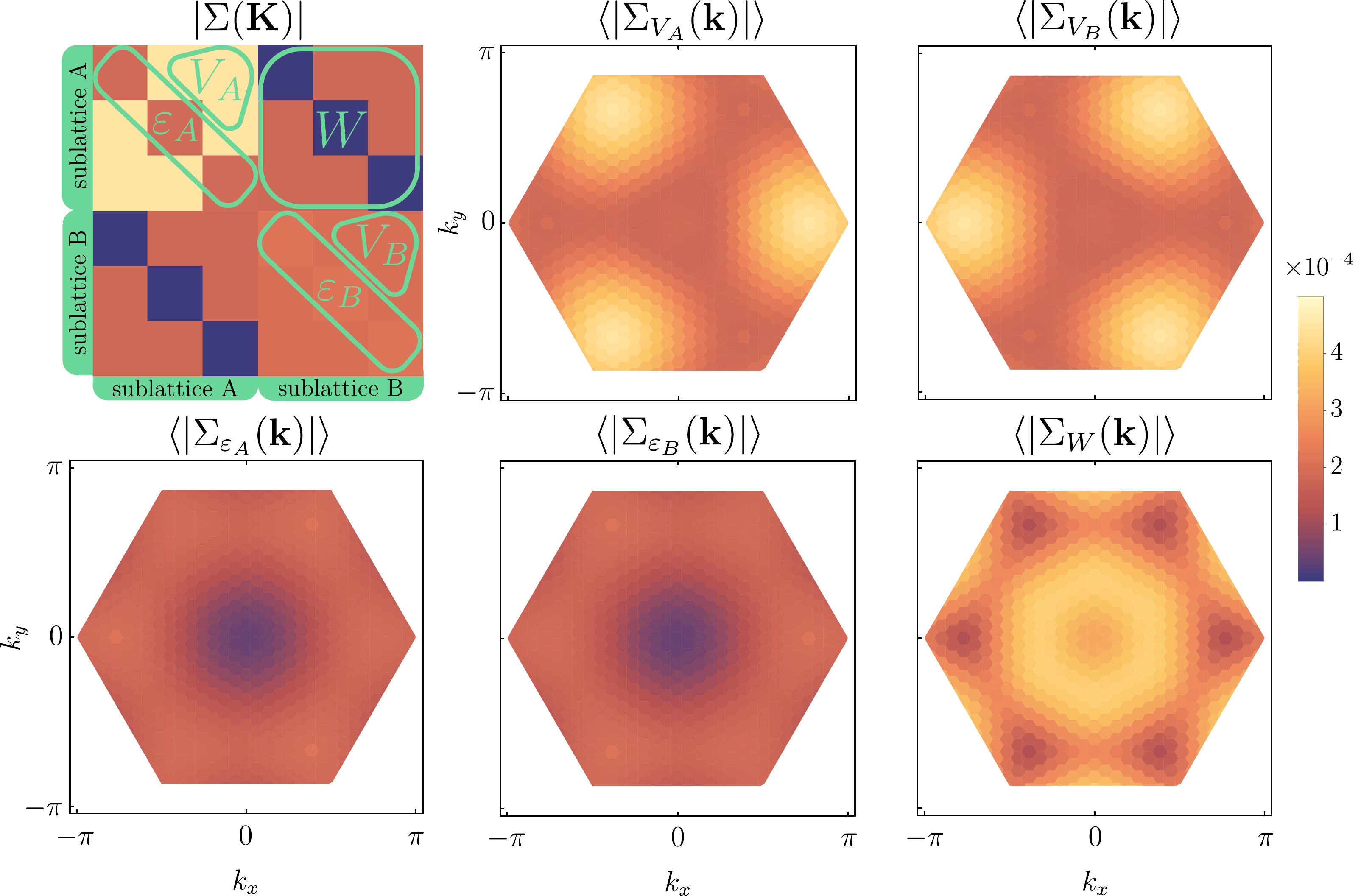}
\caption{(a)-Matrix plot of absolute values of $\Sigma(\mathbf{k})$ evaluated at $\mathbf{K}$-point, with dominant contributions given by enhanced $V$-intra-site hoppings in sublattice A. (b-g) Averages of absolute values of self-energy components across the entire BZ for $V_A,~V_B,~\varepsilon_A,~\varepsilon_B,~\text{and}~W$-terms, respectively. Intra-site hoppings $V$ show strong sublattice-valley polarization, amplifying $V_A$ hoppings for $\mathbf{K}$-valley, and $V_B$ hoppings for $\mathbf{K'}$-valley, respectively.}
\label{fig:Sigma-components}
\end{figure}

Lastly, here we discuss the distribution of self-energy components not only for the intra-site hoppings $V$, but for all components of the self-energy over the whole BZ, as a complement to the main text Fig. \ref{fig:sigma}(a,c). In Fig. \ref{fig:Sigma-components}(a) we present a matrix plot of absolute values of $\Sigma(\mathbf{K})$, as we are mainly interested in the evolution of $\mathbf{K}$-point eigenstates. Green regions encircle the corresponding elements of Bloch Hamiltonian  \ref{eq:Bloch_Hamiltonian}, namely: on-site energies $\varepsilon_A,~\varepsilon_B$, which are zero in original Hamiltonian, and hopping amplitudes $V_A,~V_B,~\text{and}~W$, which are $V_A=V_B=1-W$ in original Hamiltonian. In Fig. \ref{fig:Sigma-components}(b-g) we also show the averaged absolute values of $\Sigma(\mathbf{k})$ over the entire BZ. By "averaging" here we mean averaging over absolute values of encircled matrix elements in Fig. \ref{fig:Sigma-components}(a), for example: 
$\langle|\Sigma_{\varepsilon_A}(\mathbf{k})|\rangle = \sum^3_{i=1} |\Sigma_{ii}(\mathbf{k})|/3$.

We note that only the $V$-hopping terms obtain significant sublattice-valley polarization in the self-energy, namely the $V_A$-hopping terms are effectively amplified close to the $\mathbf{K}$ point, while the $V_B$-terms are amplified close to $\mathbf{K'}$ point. This is the main cause of the shifting of the band crossing point $\varphi_c$ to higher values of $\varphi$ in Fig.\ref{fig:sigma}(b). On the other hand, effective on-site energies get negligible sublattice-valley polarization, and $W-$hopping terms acquire local minima around both valleys.

\subsection{Geometrical properties of the effective Hamiltonian}

Being able to recover the full BZ-dependence of the self-energy, we have also studied the modification of geometric properties of the effective Hamiltonian directly in momentum space, finding how the Berry curvature and the high-symmetry eigenvalues evolve due to the presence of defects.

For this, we employ a  gauge-invariant method to compute the Berry curvature \cite{bernevig2013topological} for the band $n$:
\begin{equation}
\label{eq:berry-curv}
\Omega_n(\mathbf{k}) = -\frac{1}{2\pi}\operatorname{Im} \sum_{m \neq n} \frac{\langle n(\mathbf{k})|\nabla_{\mathbf{k}} H(\mathbf{k})|m(\mathbf{k})\rangle \times\langle m(\mathbf{k})|\nabla_{\mathbf{k}} H(\mathbf{k})|n(\mathbf{k})\rangle}{\left(E_m(\mathbf{k})-E_n(\mathbf{k})\right)^2},
\end{equation}

\begin{figure}[h!]
\includegraphics[width=0.8\textwidth]{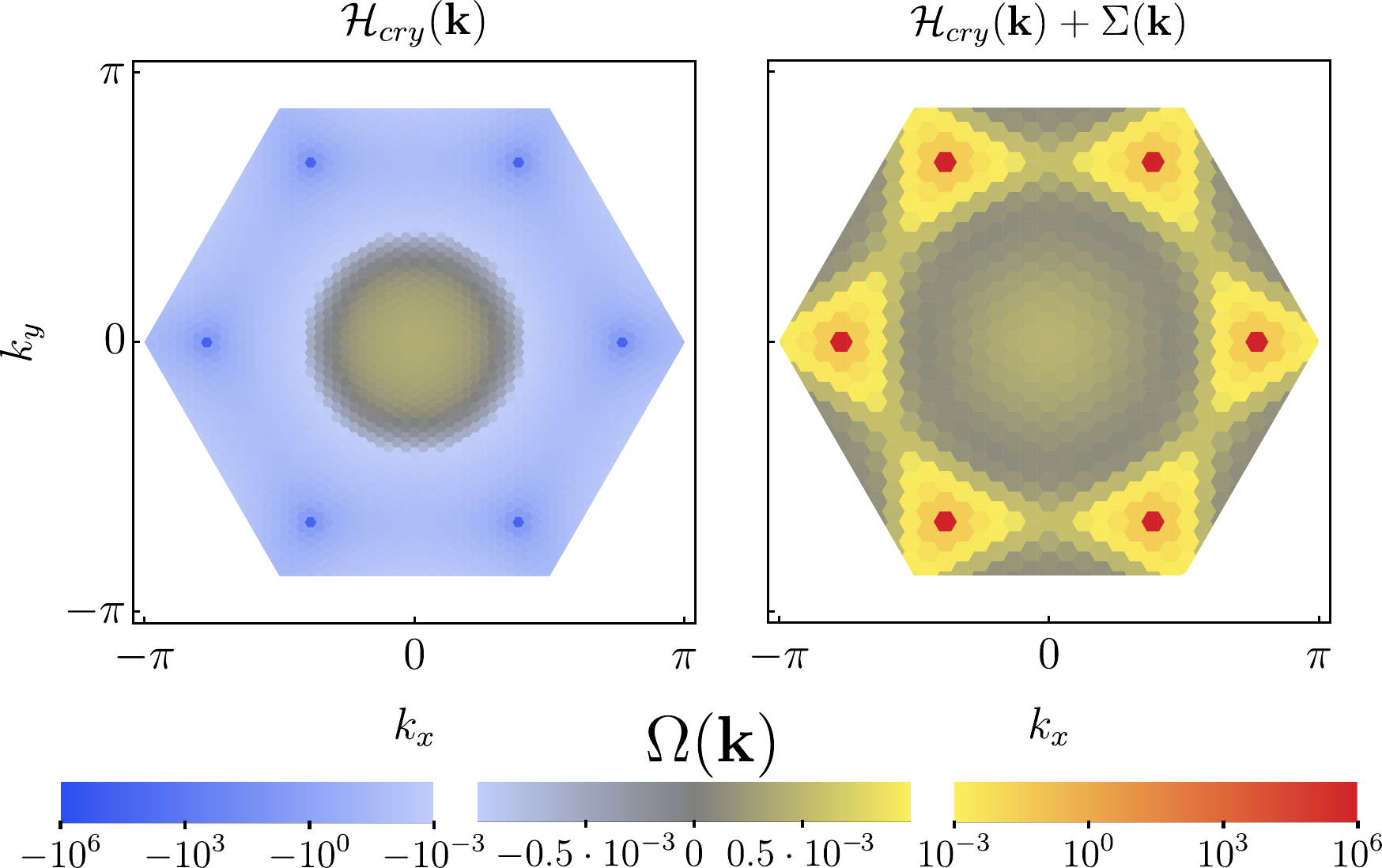}
\caption{Berry curvature distributions across the entire BZ computed by Eq.\ref{eq:berry-curv}. As the Berry curvature is strongly peaked around each valley and have very small  but finite value around $\mathbf{\Gamma}$-point, the colorscheme and legends are made in semi-logarithmic scale: (a)-Berry curvature for bare crystal $\mathcal{H}_{cry}(\mathbf{k})$ computed for finer mesh with $30\times30$ u.c. (b)-Berry curvature for effective Hamiltonian obtained from the self-energy, $\mathcal{H}_{\text{eff}}(\mathbf{k})=\mathcal{H}_{cry}(\mathbf{k})+\Sigma(\mathbf{k})$, computed for a mesh with $18\times18$ u.c. The $\Sigma(\mathbf{k})$ causes the inversion of the Berry curvature at each valley, signaling a topological phase transition with Chern number going from -1 to 1.}
\label{fig:BerryCurv}
\end{figure}

where the gradients are acting directly on the Hamiltonian instead of eigenstates. In the presence of defect averaging, however, we only have a numerical map in momentum space: $\mathcal{H}_{\text{eff}}(\mathbf{k})=\mathcal{H}_{cry}(\mathbf{k})+\Sigma(\mathbf{k})$. To find gradients $\nabla_{\mathbf{k}} \mathcal{H}_{\text{eff}}(\mathbf{k})$, we first produce a mesh of $\mathcal{H}_{cry}(\mathbf{k})+\Sigma(\mathbf{k})$ values for allowed momenta in the BZ, and then  find the gradients by computing the finite differences between neighboring mesh cells (each cell has six neighbors). The results are presented in Fig. \ref{fig:BerryCurv}. In the crystalline case (a), we use a finer mesh (30$\times$30) since we have been able to apply Eq. \ref{eq:berry-curv} directly to the Bloch Hamiltonian \ref{eq:Bloch_Hamiltonian}, which has known $\mathbf{k}$-dependence, hence no need of finite-element gradient computation. As one can see, the Berry curvature for the crystal is highly negative around each valley, and slightly positive around the $\mathbf{\Gamma}$-point. In the defect-averaged case, however, the Berry curvature changes sign close to the valleys, getting highly positive values, while curvature around $\mathbf{\Gamma}$-point remains practically intact, marking the topological phase transition, i.e., the switching of the Chern number from -1 to 1. We also emphasize the fact that the Berry curvature has a six-fold symmetry, which was recovered in $\Sigma(\mathbf{k})$ and hence in $\mathcal{H}_{\text{eff}}(\mathbf{k})$.

\begin{figure}[h!]
\includegraphics[width=0.9\textwidth]{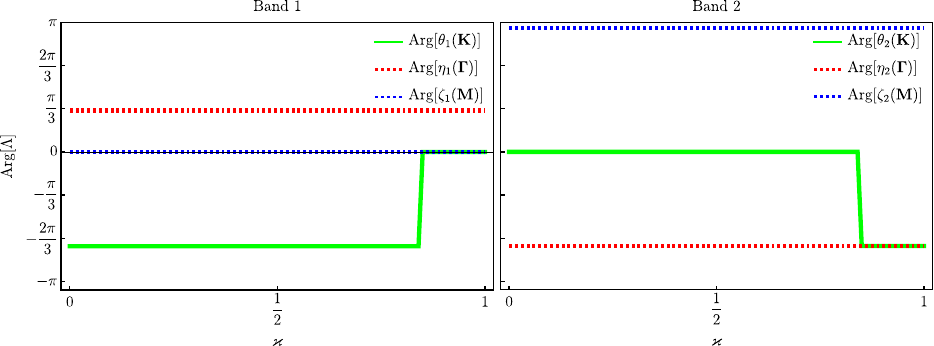}
\caption{Symmetry eigenvalues evolution for an interpolated Hamiltonian $\mathcal{H}_{cry}(\mathbf{k})+\varkappa\Sigma(\mathbf{k})$ as a function of $\varkappa$ for the two lowest-energy bands of interest: (left)-band 1, and (right)-band 2. Close to $\varkappa\approx0.85$ the symmetry eigenvalues are exchanged between the two bands, but only at the $\mathbf{K}$-point, as suggested in eq. \ref{eq:SASI}.}
\label{fig:SymmWithSigma}
\end{figure}

Since we now have access to geometrical properties of the system, and we have recovered crystalline rotational symmetries, we can make the connection with our formula in Eq. \ref{eq:SASI}. To do so, we study how the high-symmetry eigenvalues evolve due to including the self-energy. Namely, we introduce the parameter $\varkappa$, which we vary from 0 to 1 in the Hamiltonian $\mathcal{H}_{cry}(\mathbf{k})+\varkappa\Sigma(\mathbf{k})$, and at each step we find the diagonal elements of sewing matrices defined in Eq. \ref{eq:sewing_matrix_def} for each high-symmetry momentum for the two lowest-energy bands of interest. The results are presented in Fig. \ref{fig:SymmWithSigma}. As one can see, during the variation of $\varkappa$,  only the $\theta(\mathbf{K})$-eigenvalues got exchanged between the two bands, while others remained intact, which is exactly what was suggested in Eq. \ref{eq:SASI} for the Chern number computation.

\end{widetext}
\clearpage
\bibliography{main.bib}

\end{document}